\documentclass[12pt]{iopart}

\usepackage{graphicx}% Include figure files
\usepackage{dcolumn}% Align table columns on decimal point
\usepackage{bm}% bold math

\begin{document}

\title[Quantum entanglement and generalized information processing]{Quantum entanglement and generalized information processing using entangled states with
odd number of particles}

\author{Atul Kumar and Mangala Sunder Krishnan}

\address{Department of Chemistry, Indian Institute of
   Technology Madras, Chennai 600 036, India}
\ead{mangal@iitm.ac.in}
\begin{abstract}
We discuss and generalize multi-particle entanglement based on statistical correlations
using Ursell-Mayer type of cluster coefficients. Cluster coefficients are used to distinguish
different, independent entangled systems as well as those which are
connected through local unitary transformations. We propose a genuinely and
maximally entangled five-particle state for efficient information processing. The physical realization of entangled states and information processing protocols are analyzed using quantum
gates and circuit diagrams. We show that direct as
well as controlled communication can be achieved using the state
proposed here, with certainty in the case of teleportation and with a high degree of
optimity in the case of dense coding. For controlled dense coding the
amount of information transferred from the sender to the receiver is always
a maximum irrespective of the measurement basis used by the controller. 
\end{abstract}

%\submitto{XYZ Journal}
\maketitle

\section{Introduction}
Quantum entanglement is a key resource for quantum information
processing (QIP) protocols [1-4]. Information processing  involving multi-particle states
requires entangled channels which can process the information from one
remote location to another with reliability. Experimental realization of multi-particle
systems  and the detection of all orthogonal basis states
forming a complete set of entangled states remains a challenge [5-9], 
nevertheless,  efficient theoretical construction and characterization
of different multi-particle entangled channels
for analyzing different information protocols
is an important precursor to successful design of experiments.  \par
Quantum teleportation involving many particles has
been studied theoretically using
different multi-particle entangled systems [10-22]. Many experiments have
also been performed which provide partial experimental support to this
concept [23-28]. Information processing protocols such as dense coding
deal with sending classical information using an entangled quantum state as a
shared resource [29-33]. Quantum information processing techniques through nuclear magnetic resonance have been considered in detail elsewhere [34-42].\par
In this article, we propose generalized multi-particle
entangled systems for improving the efficiency of
information processing. We do this by proposing particle correlations,
as a direct measure of entanglement, using standard
Ursell-Mayer terms which are firmly founded on the principles of many body
statistical mechanics [43-47]. The approach presented here can be expanded and is applicable to
statistical  ensembles, and therefore, to electrons and other spin-1/2
systems as well as photons [48-50].  Statistical correlation
coefficients are shown to be useful in distinguishing entangled
systems belonging to different families. The properties of correlation
coefficients are used to determine whether the states under study are related through
local transformations or not. In section 3, we propose and
discuss the properties of a five-particle entangled channel and generalize the quantum channel for
$(2N+1)$ number of particles. The quantum channel proposed in that
section is used for various information processing protocols successfully. This is
followed by a conclusion.

\section{Multi-particle entanglement}
In this section, we first review the entanglement properties of a few maximally entangled
states used in the past by others and then propose multi-particle genuinely entangled
states for use in information processing. A criterion is used to
define the extent of correlation  between particles and several examples of entangled
states of many particles are considered. The entanglement properties of bipartite
states and a few multi-partite states have been studied extensively [51-60]. However, the same for
the multi-particle states is not well established. Here, the extent of entanglement
is assessed by the well established statistical mechanical formula for
correlation coefficients [43-47]. Correlation 
measures for multi-particle systems defined using Ursell-Mayer
type cluster coefficients are suggested by us as a
means for generalizing the defining of degree of entanglement between many
particles.
\subsection{Two and three-particle states} 
Correlation coefficients for two spin-1/2 particles (qubits) are defined as
%------------------------------Equation 1------------------------------------------------
\begin{eqnarray}
C^{12}_{\alpha \beta } & = & \left\langle \sigma^1_{\alpha} \sigma^2_{\beta} \right\rangle
- \left\langle \sigma ^1_{\alpha} \right\rangle
 \left\langle \sigma ^2_{\beta} \right\rangle 
\end{eqnarray}
%---------------------------------------------------------------------------------------
where $\sigma$'s are the Pauli
spin matrices for the indicated particles, 
\{$\alpha$, $\beta$, $\gamma$ = ${x, y, z}$\}. They are components of a second rank symmetric traceless tensor. 
The averages are calculated for the four Bell states of two entangled
spin-1/2 particles, namely
%----------------------------------Equation 2----------------------------------------
\begin{equation}
\left| \psi\right\rangle^{\pm}_{12}  = \frac{1}{\sqrt{2}} \left[ \,
\left| 01 \right\rangle \pm \left| 10 \right\rangle \,
\right]_{12}  ~~~{\rm and}~~~
\left| \phi\right\rangle^{\pm}_{12}  = \frac{1}{\sqrt{2}} \left[ \,
\left| 00 \right\rangle \pm \left| 11 \right\rangle \,
\right]_{12}\ 
 \end{equation}
%------------------------------------------------------------------------------------
and the non-zero correlation coefficients $(C_{xx}^{12}$,
$C_{yy}^{12}$ and  $C_{zz}^{12})$ have the absolute value $|1|$. The
maximum value  $(\pm 1)$ of correlation between the particles indicates that the states are maximally
entangled. The non-zero correlation coefficients  $(C_{xz}^{12}$,
$C_{yy}^{12}$, $C_{zx}^{12})$ for the states  
%----------------------------------Equation 3----------------------------------------
\begin{equation*}
\left| \psi'\right\rangle^{\pm}_{12}  = \frac{1}{2} \left[
\left| 00 \right\rangle-\left| 01 \right\rangle \pm \left|
  10 \right\rangle \pm \left| 11 \right\rangle
\right]_{12} ~~~{\rm and}~~~ 
\end{equation*}
\begin{equation}
\left| \phi'\right\rangle^{\pm}_{12}  = \frac{1}{2} \left[
\left| 00 \right\rangle +\left| 01 \right\rangle \pm \left|
  10 \right\rangle  \mp \left| 11 \right\rangle
\right]_{12}\ 
\end{equation}
%-------------------------------------------------------------------------------------------
which can be obtained by doing a Hadamard operation on the $2^{nd}$
particle of Bell states in Eq. (2), show
that they are also maximally entangled. The value of all correlation coefficients
associated with states such as $\left| \psi\right\rangle_{12}  = \frac{1}{2} \left[
\left| 00 \right\rangle+\left| 01 \right\rangle+\left|
  10 \right\rangle+\left| 11 \right\rangle\right]_{12}$ are zero. It is evident because $\left| \psi\right\rangle_{12}  = \frac{1}{\sqrt{2}} \left[
\left| 0\right\rangle+\left| 1 \right\rangle\right]_{1} \otimes \frac{1}{\sqrt{2}} \left[
\left| 0\right\rangle+\left| 1 \right\rangle\right]_{2}$, a direct
product state of particle 1 and particle 2.  Also the existence of the maximum value for a single
correlation coefficient alone does not ensure that a given system 
is maximally entangled, e.g. a two-particle system in a mixed state
with its density operator given by $ \rho^{12}  =
\frac{1}{2}\left[ {\left|00 \right\rangle _{12} \left\langle 00
    \right|_{12}  +  \left| {11} \right\rangle _{12} \left\langle {11}
    \right|_{12} } \right]$  shows $C_{zz}^{12}=1$, though the two particles are 
not entangled. They are nevertheless correlated in the sense that
measurement results for spin 1 and spin 2 are not independent of each
other. However, there is no ``quantum'' correlations which is due to
the off-diagonal components $\left|00 \right\rangle _{12} \left\langle 11
    \right|_{12}$ and $\left| {11} \right\rangle _{12} \left\langle {00}
    \right|_{12}$ and which is the characteristic of the entangled particles. Thus, to ensure maximum entanglement, more than one
information is needed i.e. either more than one statistical data should be
available with respect to non-zero correlation-coefficients or the
state in question must be pure along with at least one non-zero correlation
coefficient with maximum value [61]. The fact that the four Bell-states are pure and possess
more than one non-zero correlation coefficients shows that the
correlations between the particles are quantum. \par  
Correlation coefficients for the three-particle systems are
represented as 
%------------------------------Equation 4--------------------------------------------------------------------------------------
\begin{eqnarray}
C^{123}_{\alpha \beta \gamma } & = & \left\langle \sigma^1_{\alpha} \sigma^2_{\beta} \sigma^3_{\gamma}
  \right\rangle
- \left\langle \sigma ^1_{\alpha} \right\rangle
 \left\langle \sigma ^2_{\beta} \sigma^3_{\gamma} \right\rangle
- \left\langle \sigma ^2_{\beta} \right\rangle
 \left\langle \sigma ^1_{\alpha} \sigma^3_{\gamma} \right\rangle
  - \left\langle \sigma ^3_{\gamma} \right\rangle
 \left\langle \sigma ^1_{\alpha} \sigma^2_{\beta}
 \right\rangle \nonumber \\ 
&+& 2\left\langle \sigma ^1_{\alpha} \right\rangle
\left\langle \sigma ^2_{\beta} \right\rangle
\left\langle \sigma ^3_{\gamma} \right\rangle.  \nonumber \\ & &
\end{eqnarray}
%-------------------------------------------------------------------------------------------------------------------------------
They are components of a third rank tensor. The non-zero correlation coefficients $C^{123}_{\alpha \beta \gamma}$ for the three-particle
Greenberger-Horne-Zeilinger (GHZ) states [5], given by 
%----------------------------Equation 5-----------------------------------------------------------------------------------------
\begin{eqnarray}
\left| \psi \right\rangle_{123}^{(1),(2)}  = \frac{1}{\sqrt{2}} \left[ \,
\left| 000 \right\rangle \pm \left| 111 \right\rangle \,
\right]_{123} & , &
\left| \psi \right\rangle_{123}^{(3),(4)} = \frac{1}{\sqrt{2}} \left[ \,
\left| 001 \right\rangle \pm \left| 110 \right\rangle \,
\right]_{123} , \  \nonumber \\
\left| \psi \right\rangle_{123}^{(5),(6)}  =\frac{1}{\sqrt{2}} \left[ \,
\left| 010 \right\rangle \pm \left| 101 \right\rangle \,
\right]_{123} & {\rm and} &  \left| \psi \right\rangle_{234}^{(7),(8)} = \frac{1}{\sqrt{2}} \left[ \,
\left|011 \right\rangle \pm \left| 100 \right\rangle \,
\right]_{123} 
\end{eqnarray}
%------------------------------------------------------------------------------------------------------------------------------------
are  either $+1$ or $-1$ for the coefficients $(C_{xxx}^{123}$, $C_{yyx}^{123}$,
$C_{yxy}^{123}$, $C_{xyy}^{123})$. The values suggest that the correlations between
three particles are genuine and quantum.  The three-particle GHZ states, though maximally
entangled, are not robust with respect to disposal of any of the
particles i.e. tracing of any of the particles results in
the disappearance of 
quantum correlation between the rest of the particles. The other
popular three-particle entangled state is W state [62], given by 
%---------------------------------------------Eq. 6-----------------------------------------------------------------------------------------
\begin{equation}
\left|\psi\right\rangle_{123}^{W}=\frac{1}{\sqrt{3}}\left[
  \left|001\right\rangle+ \left|010\right\rangle+
  \left|100\right\rangle \right]_{123}
\end{equation}
%---------------------------------------------------------------------------------------------------------------------------------------------
have the value $\sim 4/9$ for the non-zero correlation coefficients $(C_{xxz}^{123}$,  $C_{xzx}^{123}$,  $C_{yyz}^{123}$,  $C_{yzy}^{123}$,
$C_{zxx}^{123}$,  $C_{zyy}^{123}$,  $C_{zzz}^{123})$ which suggests that the correlation between three particles is less than the maximum. The state is robust with respect to tracing of any of the particles.  A similar calculation of correlation coefficients for
a set of states such as 
%------------------Equation 7--------------------------------------------------------------------------------------------------------------------
\begin{eqnarray}
\left| \zeta \right\rangle_{123}^{(1),(2)}  &=& \frac{\left| \phi \rangle_{13}^+ \right. \otimes
\left| 0 \rangle_2 \right. \pm \left| \phi \rangle_{13}^- \right. \otimes
\left| 1 \rangle_2 \right.  }{\sqrt{2}}, \nonumber \\  
\left| \zeta \right\rangle_{123}^{(3),(4)} &=& \frac{\left| \phi \rangle_{13}^+ \right. \otimes
\left| 1 \rangle_2 \right. \pm \left| \phi \rangle_{13}^- \right. \otimes
\left| 0 \rangle_2 \right. }{\sqrt{2}}\ , \nonumber \\
\left| \zeta \right\rangle_{123}^{(5),(6)}  &=& \frac{\left| \psi \rangle_{13}^+ \right. \otimes
\left| 0 \rangle_2 \right. \pm \left| \psi\rangle_{13}^-  \right. \otimes
\left| 1 \rangle_2 \right.  }{\sqrt{2}} {\rm ~ ~ and} \nonumber \\ \left| \zeta \right\rangle_{123}^{(7),(8)} &=& \frac{\left| \psi \rangle_{13}^+\right. \otimes
\left| 1 \rangle_2 \right. \pm \left| \psi \rangle_{13}^- \right. \otimes
\left| 0 \rangle_2 \right. }{\sqrt{2}} \nonumber \\ & & 
\end{eqnarray}
%---------------------------------------------------------------------------------------------------------------------------------------
shows that these states are maximally entangled as well $(C_{xzx}^{123}$, $C_{yyx}^{123}$,
$C_{yzy}^{123}$, $C_{xyy}^{123}$ ${are \, \,  non \, \, zero})$. In
angular momentum algebraic parlance states represented in Eq. (7) and
GHZ states 
refer to different coupling schemes and can be locally transformed into
each other. The entanglement properties of these states are similar to the GHZ states if we
consider the extent of correlation between three particles. Thus, if the value of
correlation coefficients associated with a particular system is maximum
then it indicates that the state in question possesses genuine
multi-particle quantum correlations and is maximally entangled.
However, if the
value is not maximum but more than one non-zero correlation
coefficients exists the state is non-maximally entangled.
For a direct product state all the correlation
coefficients are zero suggesting no genuine multi-particle correlation
between the particles. \par 
The criteria to measure the degree of entanglement using statistical
correlations is compared with the existing criteria's such as
concurrence [52,53] (for two-particle systems) 
and with 3-tangle for three-particle maximally entangled GHZ states
and average value of square of the concurrence for less than maximally
entangled W state [54, 62]. Concurrence for a two-particle system is defined as
%-----------------------------------------Eq 8-------------------------------------------------------------------------------------------------------------
\begin{eqnarray}
C(\left|\psi\right\rangle) &=& \left\langle {\psi }\mathrel{\left | {\vphantom {\psi  {\tilde \psi }}}\right. \kern-\nulldelimiterspace}
{{\tilde \psi }} \right\rangle = \left\langle \psi \right|\sigma _y \left|\psi^{*} \right\rangle =C_{yy}^{12}.
\end{eqnarray} 
%---------------------------------------------------------------------------------------------------------------------------------------------------
where $\left| {\tilde \psi } \right\rangle  = \sigma _y \left|\psi ^{*} \right\rangle$ and $\left|\psi ^{*} \right\rangle$ is complex conjugate of $\left|\psi\right\rangle$. 
Above expression shows that the value of concurrence is equal to one of the coefficient $C_{yy}^{12}$ of second rank symmetric traceless tensor representing the correlation
between the particles. Table 1 summarizes the comparison between the
value of concurrence and correlation coefficients obtained for Bell
states. 
%------------------------------------ Table 1 --------------------------------------------------------------------------------------------------------
\begin{center}
Table 1
\end{center} 
\begin{center}
\begin{tabular}{|c|c|c|c|c|}  \hline
 state & concurrence & $C_{xx}^{12}$  & $C_{yy}^{12}$  & $C_{zz}^{12}$ \\ [2.0 ex] \hline
$\left| \psi \right\rangle_{12}^- $ & -1 & -1 & -1 & -1 \\  [2.0 ex]
$\left| \psi \right\rangle_{12}^+ $ & 1 & 1 & 1 & -1 \\  [2.0 ex] 
$\left| \phi \right\rangle_{12}^- $ & 1 & -1 & 1 & 1 \\  [2.0 ex] 
$\left| \phi\right\rangle_{12}^+ $ & -1 & 1 & -1 & 1 \\  [2.0 ex] \hline
\end{tabular}
\end{center}
%----------------------------------------------------------------------------------------------------------------------------------------------------------
The average value of the square of the 
concurrence for less than maximally entangled generalized W states
$\left|W\right\rangle_{N}$ is given by $\frac{4}{N^{2}}$. For
maximally entangled three-particle systems (ABC) such as GHZ state(s),
3-tangle is defined as
%-------------------------------Eq. 9--------------------------------------------------------------------------------------------------------------------
\begin{eqnarray}
\tau=C^{2}_{A(BC)}-C^{2}_{AB}-C^{2}_{AC}=2(\lambda_{1}^{AB}\lambda_{2}^{AB}+\lambda_{1}^{AC}\lambda_{2}^{AC}) 
\end{eqnarray} 
where $\lambda_{1}^{AB}$, $\lambda_{2}^{AB}$ and $\lambda_{1}^{AC}$, $\lambda_{2}^{AC}$ are the square roots of eigen values of $\rho ^{AB} \tilde \rho ^{AB}$ and $\rho ^{AC}
 \tilde \rho ^{AC}$, respectively such that $\tilde \rho  = (\sigma _y
 \otimes \sigma _y )\rho ^* (\sigma _y  \otimes \sigma _y )$. 
%-----------------------------------------------------------------------------------------------------------------------------------------------------
These two values are calculated and compared with that of correlation coefficients obtained using criterion used by
us. The results are summarized in Table 2 and Table 3, respectively. 
%-------------------------------------------------------------------------------------------------------------------------------- 
\begin{center}
Table 2
\end{center} 
\begin{center}
\begin{tabular}{|c|c|c|}  \hline
state & average value of square of the concurrence &
value of correlation coefficients \\ [2.0 ex] \hline
$\left|\psi\right\rangle_{123}^{W}$ & $\sim 0.45$ & $\sim 0.45$ \\   [2.0 ex] 
$\left|\psi\right\rangle_{1234}^{W}$ & $0.25$ & $0.25$ \\   [2.0 ex] 
$\left|\psi\right\rangle_{12345}^{W}$ & $0.16$ & $\sim 0.16$  \\   [2.0 ex] \hline
\end{tabular} \end{center}
%---------------------------------------------------------------------------------------------------------------------------------------------------------------
\newpage
\begin{center}
Table 3
\end{center} 
\begin{center}
\begin{tabular}{|c|c|c|c|c|c|c|c|}  \hline
state & 3-tangle & $C_{xxx}^{123}$ & $C_{yyx}^{123}$ & $C_{yxy}^{123}$ & $C_{xyy}^{123}$ & $C_{xzx}^{123}$ &
$C_{yzy}^{123}$ \\ [2.0 ex] \hline
$\frac{1}{\sqrt{2}} \left[\left| 000 \right\rangle \pm \left| 111
  \right\rangle\right]_{123}$ & 1 & $\pm 1$ & $\mp 1$ & $\mp 1$ & $\mp 1$ &
- & - \\  [2.0 ex]
$\frac{1}{\sqrt{2}} \left[\left| 001 \right\rangle \pm \left| 110
  \right\rangle\right]_{123}$ & 1 &  $\pm 1$ & $\mp 1$ & $\pm 1$ & $\pm 1$ &
- & - \\  [2.0 ex]
$\frac{1}{\sqrt{2}} \left[\left| 010 \right\rangle\pm \left| 101
  \right\rangle\right]_{123}$ & 1 &  $\pm 1$ & $\pm 1$ & $\mp 1$ & $\pm 1$ &
- & - \\  [2.0 ex] 
$\frac{1}{\sqrt{2}} \left[\left| 011 \right\rangle \pm \left| 100
  \right\rangle\right]_{123}$ & 1 &  $\pm 1$ & $\pm 1$ & $\pm 1$ & $\mp 1$ &
- & - \\  [2.0 ex] 
$\left| \zeta \right\rangle^{(1)}_{123} $ & 1 &  - & -1  &  -  &  -1 &  -1 & +1
\\  [2.0 ex] 
$\left| \zeta \right\rangle^{(2)}_{123} $ & 1 & - & -1  &  -  &  -1 &  +1 & -1
\\  [2.0 ex] 
$\left| \zeta \right\rangle^{(3)}_{123} $ & 1 & - & +1  &  -  &  +1 &  +1 & -1
\\  [2.0 ex] 

$\left| \zeta \right\rangle^{(4)}_{123} $ & 1 & - & +1  &  -  &  +1 &  -1 & +1
\\  [2.0 ex] 

$\left| \zeta \right\rangle^{(5)}_{123} $ & 1 & - & +1  &  -  &  -1 &  +1 & +1
\\  [2.0 ex] 
$\left| \chi \right\rangle^{(6)}_{123} $ & 1 & - & +1  &  -  &  -1 &  -1 & -1
\\  [2.0 ex] 

$\left| \zeta \right\rangle^{(7)}_{123} $ & 1 & - & -1  &  -  &  +1 &  -1 & -1
\\   [2.0 ex] 

$\left| \zeta \right\rangle^{(8)}_{123} $ & 1 & - & -1  &  -  &  +1 &  +1 & +1
\\ 
 [2.0 ex] \hline \end{tabular} \end{center}
%---------------------------------------------------------------------------------------------------------------------------------------------------------------------
Table 3 and Table 2 show that the value of non-zero correlation coefficients for three-particle GHZ state(s) and three-particle $\left|\zeta\right\rangle_{123}^{(i)}$ are 
in excellent argument with the value of 3-tangle whereas average value
of square of the concurrence for  $\left|W\right\rangle_{N}$ is also a
match with the value of non-zero correlation coefficients obtained. This suggests that the criterion using statistical 
correlation coefficients to measure the degree of entanglement include all possible type of entanglement in multi-particle systems and is a noble idea to study and analyze 
the properties of multi-particle systems. This can thus be generalized for arbitrary number of particles.

\subsection{Four particle systems}
The expression for four-particle correlation coefficients is given by 
%---------------------------------------------------Equation 10------------------------------------------------------------------
\begin{eqnarray} 
C^{1234}_{\alpha \beta \gamma \delta} & = & 
\left\langle \sigma^1_{\alpha} \sigma^2_{\beta} \sigma^3_{\gamma}\sigma^4_{\delta}
  \right\rangle - \left\langle \sigma ^1_{\alpha}
  \right\rangle\left[C_{\beta\gamma\delta}^{234}\right]- \left\langle
    \sigma ^2_{\beta}
  \right\rangle\left[C_{\alpha\gamma\delta}^{134}\right]- \left\langle
    \sigma ^3_{\gamma}
  \right\rangle\left[C_{\alpha\beta\delta}^{124}\right] \nonumber \\ & -& \left\langle \sigma ^4_{\delta} \right\rangle\left[C_{\alpha\beta\gamma}^{123}\right]
- \left \langle \sigma^1_{\alpha} \sigma^2_{\beta} \right \rangle 
\left \langle \sigma^3_{\gamma} \sigma^4_{\delta} \right \rangle -
 \left \langle \sigma^1_{\alpha} \sigma^3_{\gamma} \right \rangle 
\left \langle \sigma^2_{\beta} \sigma^4_{\delta} \right \rangle 
\nonumber \\ & -&  \left \langle \sigma^1_{\alpha} \sigma^4_{\delta} \right \rangle 
\left \langle \sigma^2_{\beta} \sigma^3_{\gamma} \right \rangle 
+2 \left \langle \sigma^1_{\alpha} \right \rangle \left \langle \sigma^2_{\beta} \right \rangle 
\left \langle \sigma^3_{\gamma} \right \rangle \left \langle \sigma^4_{\delta} \right \rangle  
\end{eqnarray}
%------------------------------------------------------------------------------------------------------------------------------
The non-zero correlation coefficients calculated for 
the four-particle GHZ states, namely 
%---------------------------------------------------Equation 11-----------------------------------------------------------------
\begin{equation}
\left|\psi\right\rangle_{1234}^{GHZ}=\frac{1}{\sqrt{2}}\left[\left|n_{1}n_{2}n_{3}n_{4}\right\rangle
  \pm \left|n'_{1}n'_{2}n'_{3}n'_{4}\right\rangle\right]
\end{equation}
%-----------------------------------------------------------------------------------------------------------------------------
where if $n_{i}=0$ then
$n'_{i}=1$ and {\em vice versa} are $C_{xxxx}^{1234}$, $C_{xxyy}^{1234}$, $C_{xyxy}^{1234}$,
 $C_{xyyx}^{1234}$, $C_{yxxy}^{1234}$, $C_{yxyx}^{1234}$,
 $C_{yyxx}^{1234}$, $C_{yyyy}^{1234}$, $C_{zzzz}^{1234}$ and indicate that four-particle GHZ states
possess maximum correlations. Similarly, the non-zero correlation coefficients calculated for
the four-particle W state, $\left|\psi\right\rangle_{1234}^{W}=\frac{1}{2}\left[
  \left|0001\right\rangle+ \left|0010\right\rangle +\left|0100\right\rangle+  \left|1000\right\rangle \right]_{1234}$,
are $C_{xxzz}^{1234}$, $C_{xzxz}^{1234}$, $C_{xzzx}^{1234}$,
 $C_{yyzz}^{1234}$, $C_{yzyz}^{1234}$, $C_{yzzy}^{1234}$,
 $C_{zxxz}^{1234}$, $C_{zxzx}^{1234}$, $C_{zzxx}^{1234}$,
 $C_{zyyz}^{1234}$, $C_{zyzy}^{1234}$, $C_{zzyy}^{1234}$ and 
 $C_{zzzz}^{1234}$ and show the value as $(\sim 1/4)$ indicating less
 than maximum correlations between
particles. Rigolin [17] proposed
a generalized Bell basis as a set of four-particle states to be used for
information processing, however, all the 16 four-particle correlation coefficients associated with the
generalized Bell basis are zero suggesting that there is no genuine
correlation between the four-particles. Yeo and Chua [20]
proposed a four-particle entangled system $\left| \chi
\right\rangle _{1234}^{00}$; 
the maximum value of non-zero correlation coefficients $C_{xyyx}^{1234}$,  $C_{xzzx}^{1234}$,
$C_{zyyz}^{1234}$ and  $C_{zzzz}^{1234}$ indicates that the state is
maximally and genuinely entangled. 
We consider here three sets of four-particle maximally entangled  states, in addition to GHZ states, given by
$\left|\phi\right\rangle_{1234}^{(1)-(16)}$,
$\left|\chi\right\rangle_{1234}^{(1)-(16)}$ and 
$\left|\phi'\right\rangle_{1234}^{(1)-(16)}$ where  
%---------------------------------------------Equation 12-----------------------------------------------------------------
\begin{eqnarray}
\left|\phi\right\rangle_{1234}^{(1)-(16)}
&=&\frac{1}{\sqrt{2}} \left[ \left( \begin{array}{c} \left|0\right\rangle_{1} \\ \left|1\right\rangle_{1} \\ \end{array} \right)
\otimes \left( \begin{array}{c} \left|\phi^+\right\rangle_{24} \\ \left|\psi^+\right\rangle_{24}  \\ \end{array} \right)
\otimes \left( \begin{array}{c}  \left|0\right\rangle_{3} \\
    \left|1\right\rangle_{3} \\ \end{array} \right) \right. \nonumber
\\  & & \pm
\left.\left( \begin{array}{c} \left|1\right\rangle_{1} \\ \left|0\right\rangle_{1} \\ \end{array} \right)
\otimes \left( \begin{array}{c} \left|\phi^-\right\rangle_{24}  \\ \left|\psi^-\right\rangle_{24}  \\ \end{array} \right)
\otimes \left( \begin{array}{c} \left|1\right\rangle_{3} \\
    \left|0\right\rangle_{3} \\ \end{array} \right) \right], 
\end{eqnarray}
%----------------------------------------------------------------------------------------------------------
%---------------------------------------------Equation 13-----------------------------------------------------------------
\begin{eqnarray}
\left|\chi\right\rangle_{1234}^{(1)-(16)}
&=&\frac{1}{\sqrt{2}} \left[ \left( \begin{array}{c} \left|0\right\rangle_{1} \\ \left|1\right\rangle_{1} \\ \end{array} \right)
\otimes \left( \begin{array}{c} \left|\phi^+\right\rangle_{24} \\ \left|\phi^-\right\rangle_{24}  \\ \end{array} \right)
\otimes \left( \begin{array}{c}  \left|0\right\rangle_{3} \\
    \left|1\right\rangle_{3} \\ \end{array} \right) \right. \nonumber
\\  & & \pm
\left.\left( \begin{array}{c} \left|1\right\rangle_{1} \\ \left|0\right\rangle_{1} \\ \end{array} \right)
\otimes \left( \begin{array}{c} \left|\psi^-\right\rangle_{24}  \\ \left|\psi^+\right\rangle_{24}  \\ \end{array} \right)
\otimes \left( \begin{array}{c} \left|1\right\rangle_{3} \\
    \left|0\right\rangle_{3} \\ \end{array} \right) \right], 
\end{eqnarray}
%----------------------------------------------------------------------------------------------------------
and
%---------------------------------------------Equation 14-----------------------------------------------------------------
\begin{eqnarray}
\left|\phi'\right\rangle_{1234}^{(1)-(16)}
&=&\frac{1}{\sqrt{2}} \left[ \left( \begin{array}{c} \left|0\right\rangle_{1} \\ \left|1\right\rangle_{1} \\ \end{array} \right)
\otimes \left( \begin{array}{c} \left|\phi^+\right\rangle_{24} \\ \left|\phi^-\right\rangle_{24}  \\ \end{array} \right)
\otimes \left( \begin{array}{c}  \left|0\right\rangle_{3} \\
    \left|1\right\rangle_{3} \\ \end{array} \right) \right. \nonumber
\\  & & \pm
\left.\left( \begin{array}{c} \left|1\right\rangle_{1} \\ \left|0\right\rangle_{1} \\ \end{array} \right)
\otimes \left( \begin{array}{c} \left|\psi^+\right\rangle_{24}  \\ \left|\psi^-\right\rangle_{24}  \\ \end{array} \right)
\otimes \left( \begin{array}{c} \left|1\right\rangle_{3} \\
    \left|0\right\rangle_{3} \\ \end{array} \right) \right].
\end{eqnarray}
%----------------------------------------------------------------------------------------------------------
The non-zero correlation coefficients calculated for the above
three sets are $(C_{xxyy}^{1234}$, $C_{xyyx}^{1234}$, $C_{yxxy}^{1234}$ and
$C_{yyxx}^{1234})$, $(C_{xxxz}^{1234}$, $C_{xzxx}^{1234}$, $C_{yxyz}^{1234}$,
$C_{yzyx}^{1234})$ and $(C_{xzyy}^{1234}$, $C_{xyyz}^{1234}$, $C_{yzxy}^{1234}$,
$C_{yyxz}^{1234})$, respectively and indicate maximum entanglement. The set of states represented by Eq. (12) and
Eq. (14) are cluster type of states [63] and can be transformed into each other through local transformations whereas Eq. (13) represents $\left|\chi\right\rangle$ type of states [20] .
\subsection{Five-particle systems}
The expression for the five-particle correlation coefficient is given
in the Appendix A. The generalized five-particle GHZ states are represented as
%---------------------------------------------------Equation 13-----------------------------------------------------------------
\begin{equation}
\left|\psi\right\rangle_{12345}^{GHZ}=\frac{1}{\sqrt{2}}\left[\left|n_{1}n_{2}n_{3}n_{4}n_{5}\right\rangle
  \pm \left|n'_{1}n'_{2}n'_{3}n'_{4}n'_{5}\right\rangle\right]
\end{equation}
%-----------------------------------------------------------------------------------------------------------------------------
and are maximally
correlated as shown by non-zero correlation coefficients $C^{12345}_{xxxxx}$, $C^{12345}_{xxxyy}$,
$C^{12345}_{xxyxy}$, $C^{12345}_{xxyyx}$, $C^{12345}_{xyxxy}$, $C^{12345}_{xyxyy}$,
$C^{12345}_{xyyxx}$, $C^{12345}_{xyyyy}$, $C^{12345}_{yxxxy}$,
$C^{12345}_{yxxyx}$, $C^{12345}_{yxyxx}$, $C^{12345}_{yxyyy}$,
$C^{12345}_{yyxxx}$, $C^{12345}_{yyxyy}$, $C^{12345}_{yyyxy}$ and
$C^{12345}_{yyyyx}$. Unlike the GHZ state, the
generalized five-particle W state, $\left|\psi\right\rangle_{12345}^{W}=\frac{1}{\sqrt{5}}\left[
  \left|00001\right\rangle+ \left|00010\right\rangle+
  \left|00100\right\rangle+  \left|01000\right\rangle \right. \nonumber
\\ +
  \left. \left|10000\right\rangle
\right]_{12345}$, is not maximally correlated as shown by non-zero correlation coefficients $C^{12345}_{xxzzz}$, $C^{12345}_{xzxzz}$,
$C^{12345}_{xzzxz}$, $C^{12345}_{xzzzx}$, $C^{12345}_{yyzzz}$, $C^{12345}_{yzyzz}$,
$C^{12345}_{yzzyz}$, $C^{12345}_{yzzzy}$, $C^{12345}_{zxxzz}$,
$C^{12345}_{zxzxz}$, $C^{12345}_{zxzzx}$, $C^{12345}_{zyyzz}$,
$C^{12345}_{zyzyz}$, $C^{12345}_{zyzzy}$, $C^{12345}_{zzxxz}$, $C^{12345}_{zzxzx}$,
$C^{12345}_{zzyyz}$, $C^{12345}_{zzyzy}$, $C^{12345}_{zzzxx}$,
$C^{12345}_{zzzyy}$ and $C^{12345}_{zzzzz}$. Other five-particle entangled systems to be considered are two sets of basis states given as
$\left|\Psi\right\rangle_{12345}^{(1)-(32)}$ and
  $\left|\Phi\right\rangle_{12345}^{(1)-(32)}$, where  
%---------------------------------Equation 14 and 15------------------------------------------------------------------------------------------------
\begin{eqnarray}
\left|\Psi\right\rangle_{12345}^{(1)-{(32)}}&=&
 \frac{1}{\sqrt{2}} \left[ \left( \begin{array}{c} \left|0\right\rangle_{1} \\ \left|1\right\rangle_{1} \\ \end{array} \right)
\otimes \left( \begin{array}{c} \left|\psi^{(1)}\right\rangle_{234} \\
    \left|\psi^{(2)}\right\rangle_{234}  \\
    \left|\psi^{(3)}\right\rangle_{234} \\ 
  \left|\psi^{(4)}\right\rangle_{234} \\ \end{array} \right)
\otimes \left( \begin{array}{c}  \left|0\right\rangle_{5} \\
    \left|1\right\rangle_{5} \\ \end{array} \right) \right. \nonumber
\\  & & \pm 
\left. \left( \begin{array}{c} \left|1\right\rangle_{1} \\ \left|0\right\rangle_{1} \\ \end{array} \right)
\otimes \left( \begin{array}{c} \left|\psi^{(6)}\right\rangle_{234}
    \\\left|\psi^{(5)}\right\rangle_{234}  \\
    \left|\psi^{(8)}\right\rangle_{234} \\
    \left|\psi^{(7)}\right\rangle_{234} \\ \end{array} \right)
\otimes \left( \begin{array}{c} \left|1\right\rangle_{5} \\
    \left|0\right\rangle_{5} \\ \end{array} \right) \right]
\end{eqnarray} 
%------------------------------------------------------------------------------------------------------------------------------------
{\rm ~ ~ ~ and}
\begin{eqnarray}
\left|\Phi\right\rangle_{12345}^{(1)-{(32)}}&=&
\frac{1}{\sqrt{2}} \left[ \left( \begin{array}{c} \left|0\right\rangle_{1} \\ \left|1\right\rangle_{1} \\ \end{array} \right)
\otimes \left( \begin{array}{c} \left|\psi^{(1)}\right\rangle_{234} \\
    \left|\psi^{(2)}\right\rangle_{234}  \\
    \left|\psi^{(3)}\right\rangle_{234} \\ 
  \left|\psi^{(4)}\right\rangle_{234} \\ \end{array} \right)
\otimes \left( \begin{array}{c}  \left|0\right\rangle_{5} \\
    \left|1\right\rangle_{5} \\ \end{array} \right) \right. \nonumber
\\ & &  \pm
\left. \left( \begin{array}{c} \left|1\right\rangle_{1} \\ \left|0\right\rangle_{1} \\ \end{array} \right)
\otimes \left( \begin{array}{c} \left|\psi^{(5)}\right\rangle_{234}
    \\\left|\psi^{(6)}\right\rangle_{234}  \\
    \left|\psi^{(7)}\right\rangle_{234} \\
    \left|\psi^{(8)}\right\rangle_{234} \\ \end{array} \right)
\otimes \left( \begin{array}{c} \left|1\right\rangle_{5} \\
    \left|0\right\rangle_{5} \\ \end{array} \right) \right]. 
\end{eqnarray}
%-----------------------------------------------------------------------------------------------------------------------------------
$\left|\psi\right\rangle_{234}^{(1)-(8)}$ are three-particle GHZ states and are given by Eq. (5). The non-zero correlation coefficients for the two
sets are $C^{12345}_{xxzxx}$, $C^{12345}_{xyzyx}$, $C^{12345}_{yxzxy}$,
$C^{12345}_{yyzyy}$ and  
$C^{12345}_{xxzyy}$, $C^{12345}_{xyzxy}$, $C^{12345}_{yxzyx}$,
$C^{12345}_{yyzxx}$, respectively and show maximum value. The extent of correlation between
five-particles remains the same even after interchanging the particle
indices. \par
The general expression for the $N$-particle correlation coefficient can be obtained 
by solving the equations for cluster functions derived formally from
 the $N$-th quantum virial coefficient.
The following summarizes the relation between correlation
coefficients and the degree of entanglement. \\  
(i) Existence of maximum values for more than one correlation coefficient for a system under
study, indicates that the state of the system possesses genuine and maximum
entanglement. \\
(ii) For non-maximally entangled states the value of correlation
coefficients lies between $0$ and $1$. \\ 
(iii) Null results for all correlation coefficients of state suggest
that it is a direct product of fewer particle states and there
exists no genuine multi-particle entanglement. \\
(iv) The value of non-zero correlation coefficients remains the same for states 
connected to each other by local unitary transformations. \\
(v) The extent of correlation remains invariant to
changing the particle indices. \par
\subsection{Importance and properties of cluster coefficients}
The criterion to use cluster coefficients as a measure of 
entanglement of the state under study allows one to
characterize the extent of correlation of multi-particle states on the
same scale irrespective of number of particles involved. A consistent description emerges for systems
irrespective of the number of particles which are entangled. In this subsection, we
discuss some of the  properties of correlation coefficients in addition to those described in
previous section.
\begin{enumerate}
\item The relation between correlation coefficients of states which
      differ from each other only through permutation of particle
      indices can be seen immediately as follows: \\
(1)  The state $\left|\phi\right\rangle_{1234}^{(1)}=\frac{1}{2}\left[\left|0000\right\rangle
+ \left|0101\right\rangle + \left|1010\right\rangle -
\left|1111\right\rangle\right]_{1234}$ [Eq. (12)] is obtained from $\left|\phi''\right\rangle_{1234}^{(1)}=\frac{1}{2}\left[\left|0000\right\rangle
+ \left|1001\right\rangle + \left|0110\right\rangle -
\left|1111\right\rangle\right]_{1234}$ by permuting particles 1 and
2. Hence, the non-zero correlation coefficients
associated with $\left|\phi\right\rangle_{1234}^{(1)}$ and
$\left|\phi''\right\rangle_{1234}^{(1)}$ are $(C_{xxyy}^{1234}$, $C_{xyyx}^{1234}$, $C_{yxxy}^{1234}$,
$C_{yyxx}^{1234})$ and $(C_{xxyy}^{1234}$, $C_{xyxy}^{1234}$, $C_{yxyx}^{1234}$,
$C_{yyxx}^{1234})$, respectively.  \\
(2) Conversely, by examining two sets of equal number of correlation
coefficients, we can also relate the states. For example, the two sets
$(C^{12345}_{xxzxx}$, $C^{12345}_{xyzyx}$, $C^{12345}_{yxzxy}$,
$C^{12345}_{yyzyy})$ and $(C^{12345}_{xxxxz}$, $C^{12345}_{yxxyz}$, $C^{12345}_{xyyxz}$,
$C^{12345}_{yyyyz})$ are related to each other through particle
permutations $(1 \leftrightarrow 2)$ and  $(3 \leftrightarrow 5)$. The
first set is the only non-null set of coefficients for the state given
by Eq. (16). Hence another can be obtained by particle
permutations. Thus, a family of states can be quickly enumerated.  

\item If the number of non-zero correlation coefficients corresponding to
      two entangled sets are not equal, then they belong to two different
      family of states. \\
(1) Three-particle GHZ state $(C_{xxx}^{123}$, $C_{yyx}^{123}$,
$C_{yxy}^{123}$, $C_{xyy}^{123})$ and three-particle W state
$(C_{xxz}^{123}$, $C_{xzx}^{123}$, $C_{yyz}^{123}$, $C_{yzy}^{123}$,
$C_{zxx}^{123}$,  $C_{zyy}^{123}$,  $C_{zzz}^{123})$ show four
and seven non-zero correlation coefficients, respectively. They belong
to two different families of states. \\
(2) Four-particle maximally entangled GHZ states $(C_{xxxx}^{1234}$, $C_{xxyy}^{1234}$, $C_{xyxy}^{1234}$, $C_{xyyx}^{1234}$,
$C_{yxxy}^{1234}$, $C_{yxyx}^{1234}$, $C_{yyxx}^{1234}$, $C_{yyyy}^{1234}$,
$C_{zzzz}^{1234})$ represented by
Eq. (11) and four-particle maximally entangled set represented by
Eq. (12) $(C_{xxyy}^{1234}$, $C_{yxxy}^{1234}$, $C_{xyyx}^{1234}$,
$C_{yyxx}^{1234})$ show nine and four non-zero correlation coefficients, respectively 
which indicates that these two sets belong to different families of
states. \\
(3) The set of five-particle states represented by Eq. (17) $(C^{12345}_{xxzyy}$, $C^{12345}_{xyzxy}$, $C^{12345}_{yxzyx}$,
$C^{12345}_{yyzxx})$ and five-particle Brown state [16] $(C^{12345}_{xxyyx}$,
$C^{12345}_{xxzxz}$, $C^{12345}_{yyzzx}$, $C^{12345}_{yzxxy}$, \\
$C^{12345}_{zxyzy}$, $C^{12345}_{zyxyz})$ possess four
and six non-zero correlation coefficients, respectively and hence belong to two  
different families of entangled systems. 
\item Even if the number of correlation coefficients associated with
      different entangled states are equal the states need not have to
      belong to the same family. For example, the three states 
$\left|\phi\right\rangle_{1234}^{(1)}=\frac{1}{2}\left[\left|0000\right\rangle
+ \left|0101\right\rangle + \left|1010\right\rangle -
\left|1111\right\rangle\right]_{1234}$, 
$\left|\chi\right\rangle_{1234}^{(1)}=\frac{1}{2}\left[\left|0000\right\rangle
+ \left|0101\right\rangle + \left|1011\right\rangle -
\left|1110\right\rangle\right]_{1234}$ and
$\left|\phi'\right\rangle_{1234}^{(1)}=\frac{1}{2}\left[\left|0000\right\rangle
+ \left|0101\right\rangle \right. \nonumber \\ 
  +\left. \left|1011\right\rangle+ \left|1110\right\rangle\right]_{1234}$ belong to maximally entangled four-particle sets
represented by Eq. (12), Eq. (13) and Eq. (14), respectively and
possess four non-zero correlation coefficients. Although $\left|\phi\right\rangle_{1234}^{(1)}$  and
$\left|\phi'\right\rangle_{1234}^{(1)}$  belong to same family,
$\left|\chi\right\rangle_{1234}^{(1)}$  belongs to different family of
states. 
\item The extent of correlation between particles remains
      invariant under standard local unitary transformations. The set of states $(\left|\zeta\right\rangle_{123}^{(i)})$ represented
  by Eq. (7) can be obtained by
applying Hadamard operation to the second
particle of three-particle GHZ states (Eq. (5)) and possesses the same
degree of correlation as that of GHZ states.
\item Doing a Hadamard operation would
      not affect the y-component of the correlation coefficient but  
      would convert x-component to z-component and {\em vice
        versa}. Thus, the non-zero correlation coefficients for the set of states
represented by Eq. (2) and Eq. (3) (which differ from each other by a
Hadamard transformation on particle 2) are $(C_{xx}^{12}$, $C_{yy}^{12}$,
$C_{zz}^{12})$ and $(C_{xz}^{12}$, $C_{yy}^{12}$,
$C_{zx}^{12})$, respectively. This is a trivial example. However, as the
number of particles in an entangled set increases, the number of ways of doing
transformations also increases and hence this scheme is useful for
nontrivial, multiple local transformations as shown below. \\
(1) The correlation coefficients,  $(C_{xyyx}^{1234}$, $C_{xzzx}^{1234}$, $C_{zyyz}^{1234}$,
$C_{zzzz}^{1234})$ and  $(C_{yxyz}^{1234}$, $C_{xxxz}^{1234}$, $C_{yzyx}^{1234}$,
$C_{xzxx}^{1234})$, correspond to the entangled states $\left| \chi \right\rangle _{1234}^{00} =  \frac
{1}{2\sqrt{2}} \left[ \left|0000\right\rangle-\left|0011\right\rangle
\right. \nonumber \\ 
  - \left.\left|0101\right\rangle+\left|0110\right\rangle
  +\left|1001\right\rangle+\left|1010\right\rangle
  +\left|1100\right\rangle
  +\left|1111\right\rangle\right]_{1234}$
and $\left|\chi\right\rangle_{1234}^{(1)}=\frac{1}{2}\left[\left|0000\right\rangle\right. \nonumber \\ 
  +\left.\left|0101\right\rangle +\left|1011\right\rangle -
\left|1110\right\rangle\right]_{1234}$, respectively. The set of
coefficients can be transformed into each other by doing Hadamard
transformations on $2^{nd}$, $3^{rd}$ and $4^{th}$ particles and
permuting particles 1 and 2. \\ 
(2) The five-particle maximally entangled states, namely  $\left|\varsigma\right\rangle_{12345}^{(1)}=\frac{1}{2}\left[\left|00000\right\rangle
\right. \nonumber \\ 
  + \left. \left|01110\right\rangle+\left|10001\right\rangle -
\left|11111\right\rangle\right]_{12345}$ and
$\left|\varsigma'\right\rangle _{12345}^{(1)} =  \frac
{1}{2\sqrt{2}} \left[ \left|00000\right\rangle+\left|00110\right\rangle\right. \nonumber \\ 
  + \left. \left|01010\right\rangle +\left|01100\right\rangle + \left|10011\right\rangle+\left|10101\right\rangle
  +\left|11001\right\rangle
  +\left|11111\right\rangle\right]_{12345}$ can be
converted into one another by doing local transformations as revealed
by their correlation coefficients, namely  
$(C^{12345}_{xxxyy}$,
$C^{12345}_{xxyxy}$, $C^{12345}_{xyxxy}$, $C^{12345}_{xyyyy}$,
$C^{12345}_{xzzzx}$, $C^{12345}_{yxxyx}$, $C^{12345}_{yxyxx}$,
$C^{12345}_{yyxxx}$, $C^{12345}_{yyyyx}$, $C^{12345}_{yzzzy})$ and $(C^{12345}_{xzzyy}$,
$C^{12345}_{xzyzy}$, $C^{12345}_{xyzzy}$, $C^{12345}_{xyyyy}$,
$C^{12345}_{xxxxx}$, $C^{12345}_{yzzyx}$, $C^{12345}_{yzyzx}$,
$C^{12345}_{yyzzx}$, $C^{12345}_{yyyyx}$, $C^{12345}_{yxxxy})$, 
respectively. Thus by doing three Hadamard operations on particle two, three and
four $\left|\varsigma\right\rangle_{12345}^{(1)}$
can be locally transformed to
$\left|\varsigma'\right\rangle_{12345}^{(1)}$. 
\end{enumerate}

\section{Generalized information processing}
In this section we propose a maximally and genuinely entangled five-particle state and describe different information processing protocols
using the state. In the past, multi-particle entangled
channels involving odd number of particles have been proposed {\em with the
use of a controller} to assist the sender for successful and optimal
information transfer [10, 13, 14]. We show that one
can {\em eliminate the intermediate observer} controlling the process such that information
processing is successful in all the measurement outcomes performed by
the sender. The formation of the state proposed here ensures efficient information transfer between two or more users in the communication protocol. 
\subsection{Direct teleportation}
The five-particle maximally entangled set proposed here is
given by 
%-----------------------Equation 16----------------------------------------------------------------------------------------
\begin{eqnarray}
\lefteqn{\left|\varphi\right\rangle_{12345}^{(1)-(32)} =} & & \nonumber \\ & & 
\frac{1}{\sqrt{2}} \left[  \left( \begin{array}{c} \left|\chi^{(1)}\right\rangle_{1234} \\
    \left|\chi^{(2)}\right\rangle_{1234}  \\
    \left|\chi^{(3)}\right\rangle_{1234} \\ 
  \left|\chi^{(4)}\right\rangle_{1234} \\  \left|\chi^{(5)}\right\rangle_{1234}  \\
    \left|\chi^{(6)}\right\rangle_{1234} \\
    \left|\chi^{(7)}\right\rangle_{1234} \\
    \left|\chi^{(8)}\right\rangle_{1234} \\ \end{array} \right)
\otimes \left( \begin{array}{c}  \left|0\right\rangle_{5} \\  \left|1\right\rangle_{5} \\ \end{array} \right)  \pm
 \left( \begin{array}{c} \left|\chi^{(9)}\right\rangle_{1234}
   \\ \left|\chi^{(10)}\right\rangle_{1234}  \\
    \left|\chi^{(11)}\right\rangle_{1234} \\
    \left|\chi^{(12)}\right\rangle_{1234} \\  \left|\chi^{(13)}\right\rangle_{1234}  \\
    \left|\chi^{(14)}\right\rangle_{1234} \\
    \left|\chi^{(15)}\right\rangle_{1234} \\
    \left|\chi^{(16)}\right\rangle_{1234} \\ \end{array} \right)
\otimes \left( \begin{array}{c} \left|1\right\rangle_{5} \\
    \left|0\right\rangle_{5} \\ \end{array} \right) \right] \nonumber \\ & & 
\end{eqnarray}
%-------------------------------------------------------------------------------------------------------------------------------------
where $\left|\chi\right\rangle_{1234}^{(1)-(16)}$
are given by 
%---------------------------------------------Equation 17-----------------------------------------------------------------
\begin{eqnarray}
\left| \chi \right\rangle_{1234}^{(1),(2)} & =  & \frac{\left| 0 \rangle_1
  \right. \otimes \left| \phi \rangle_{24}^+ \right. \otimes
\left| 0 \rangle_3 \right. \pm \left| 1 \rangle_1
  \right. \otimes \left| \psi \rangle_{24}^- \right. \otimes
\left| 1 \rangle_3 \right.  }{\sqrt{2}} ,  \nonumber \\ 
\left| \chi \right\rangle_{1234}^{(3),(4)} & = & \frac{\left| 0 \rangle_1
  \right. \otimes \left| \phi \rangle_{24}^- \right. \otimes
\left| 0 \rangle_3 \right. \pm \left| 1 \rangle_1
  \right. \otimes \left| \psi \rangle_{24}^+ \right. \otimes
\left| 1 \rangle_3 \right.  }{\sqrt{2}} ,  \nonumber \\
\left| \chi \right\rangle_{1234}^{(5),(6)} & =  & \frac{\left| 1 \rangle_1
  \right. \otimes \left| \phi \rangle_{24}^+ \right. \otimes
\left| 0 \rangle_3 \right. \pm \left| 0 \rangle_1
  \right. \otimes \left| \psi \rangle_{24}^- \right. \otimes
\left| 1 \rangle_3 \right.  }{\sqrt{2}} ,  \nonumber \\ 
\left| \chi \right\rangle_{1234}^{(7),(8)} & = & \frac{\left| 1 \rangle_1
  \right. \otimes \left| \phi \rangle_{24}^- \right. \otimes
\left| 0 \rangle_3 \right. \pm \left| 0 \rangle_1
  \right. \otimes \left| \psi \rangle_{24}^+ \right. \otimes
\left| 1 \rangle_3 \right.  }{\sqrt{2}} ,  \nonumber \\
\left| \chi \right\rangle_{1234}^{(9),(10)}& =  & \frac{\left| 0 \rangle_1
  \right. \otimes \left| \phi \rangle_{24}^+ \right. \otimes
\left| 1 \rangle_3 \right. \mp \left| 1 \rangle_1
  \right. \otimes \left| \psi \rangle_{24}^- \right. \otimes
\left| 0 \rangle_3 \right.  }{\sqrt{2}} , \nonumber \\ 
\left| \chi \right\rangle_{1234}^{(11),(12)} & =  & \frac{\left| 0 \rangle_1
  \right. \otimes \left| \phi \rangle_{24}^- \right. \otimes
\left| 1 \rangle_3 \right. \mp \left| 1 \rangle_1
  \right. \otimes \left| \psi \rangle_{24}^+ \right. \otimes
\left| 0 \rangle_3 \right.  }{\sqrt{2}} , \nonumber \\
\left| \chi\right\rangle_{1234}^{(13),14)} & =  & \frac{\left| 1 \rangle_1
  \right. \otimes \left| \phi \rangle_{24}^+ \right. \otimes
\left| 1 \rangle_3 \right. \mp \left| 0 \rangle_1
  \right. \otimes \left| \psi \rangle_{24}^- \right. \otimes
\left| 0 \rangle_3 \right.  }{\sqrt{2}} {\rm ~ ~ and} \nonumber \\
\left| \chi \right\rangle_{1234}^{(15),(16)} & =  & \frac{\left| 1 \rangle_1
  \right. \otimes \left| \phi \rangle_{24}^- \right. \otimes
\left| 1 \rangle_3 \right. \mp \left| 0 \rangle_1
  \right. \otimes \left| \psi \rangle_{24}^+ \right. \otimes
\left| 0 \rangle_3 \right.  }{\sqrt{2}}.
\end{eqnarray}
%----------------------------------------------------------------------------------------------------------
The set represented above is same as four-particle entangled set given in Eq. (13), however, the order in which the states are represented is different.  The set proposed here shows values $\pm 1$ for the non-zero correlation coefficients $(C_{xxxzz}^{12345}$, $C_{xxzzx}^{12345}$, $C_{xzxxz}^{12345}$,
$C_{xzzxx}^{12345})$. Depending on the discussions of previous section and due to the absence of 'y' in
the $C$'s the five-particle entangled set proposed above 
belongs to a different family of states with respect to those represented by
Eq. (16) and Eq. (17). \par
In order to communicate an arbitrary two-particle information to Bob i.e.  $\left|\phi\right\rangle_{12} = 
\left[a\left|00\right\rangle+b\left|01\right\rangle
+c\left|10\right\rangle+d\left|11\right\rangle\right]_{12}$, Alice must
share any one of the five-particle entangled state
$\left|\varphi\right\rangle_{34567}^{(i)}$ given by Eq. (18) with 
Bob such that particles 3, 4 and 5 are with Alice  and particles 6
and 7 are with Bob. Thus, using 
$\left|\varphi\right\rangle_{34567}^{(10)}$ as the quantum channel
shared between Alice and Bob where
$\left|\varphi\right\rangle_{34567}^{(10)}=\frac{1}{2\sqrt{2}}\left[\left|00000\right\rangle-\left|00101\right\rangle\right. \nonumber \\ 
  + \left.
\left|11100\right\rangle+\left|11001\right\rangle+\left|01111\right\rangle-\left|01010\right\rangle+\left|10011\right\rangle+\left|10110\right\rangle\right]_{34567}$,
Alice can communicate her unknown message with Bob by interacting her
particles 1 and 2 with her share of entangled particles 3, 4 and 5 so that
%------------------------------Equation 18---------------------------------------------------------------------------------------------
\begin{eqnarray}
\left|\psi\right\rangle_{1234567} &=&
\left|\phi\right\rangle_{12}\otimes\left|\varphi\right\rangle_{34567}^{(10)}. 
\end{eqnarray}
%------------------------------------------------------------------------------------------------------------------------------------------------------------
Eq. (20) can be re-expressed in form of Alice's projection basis given
by Eq. (17) as 
%------------------------------Equation 19----------------------------------------------------------------------------------------------
\begin{eqnarray}
\left|\psi\right\rangle_{1234567}=\frac{1}{4\sqrt{2}}\sum_{i,j}\left|\Phi\right\rangle^{(i)}_{12345}\otimes\left|\phi\right\rangle_{67}^{(j)}
\end{eqnarray}
%-----------------------------------------------------------------------------------------------------------------------------------------------------------
where $i=1, 32$ and $j=1, 4$. For Alice's
measurement outcomes $\left|\Phi\right\rangle^{(1)}_{12345}$ and
$\left|\Phi\right\rangle^{(30)}_{12345}$, Bob's particles are
instantaneously projected on to the state Alice wanted to communicate,
with total probability of 1/16, however, for all other measurement
outcomes of Alice, Bob require  {\em only} single qubit transformations to recover the message
successfully. The preparation of above set of states and teleportation of an arbitrary two-particle state are represented in Fig. (1) and Fig. (2), respectively.
\subsection{Controlled teleportation}
For controlled teleportation, the quantum state
$\left|\varphi\right\rangle_{34567}^{(10)}$ is shared between Alice, Charlie and Bob such that the particles 3 and 4 are with
Alice, particles 5 and 6 are with Bob and particle 7 is with 
Charlie. Alice projects her four particles on to the basis set given
by Eq. (13) so that Eq. (20) becomes
%----------------------------Equation 20-------------------------------------------------------------------------------------
\begin{eqnarray}
\left|\psi\right\rangle_{1234567}=\frac{1}{4}\sum_{i,j}
\left|\chi\right\rangle_{1234}^{(i)}\otimes\left|\psi\right\rangle_{567}^{(j)}
\end{eqnarray}  
%---------------------------------------------------------------------------------------------------------------------------------------------------------
where $i=1, 16$ and $j=1, 8$. For example, if
Alice's measurement outcome is
$\left|\chi\right\rangle_{1234}^{(5)}$, the combined state of Bob's
and Charlie's particles is given by 
%----------------------------Equation 21------------------------------------------------------------------------------------
\begin{eqnarray}
\left|\psi\right\rangle_{567} &=& \frac{1}{\sqrt{2}}\left
  [a\left|00\right\rangle_{56}+b\left|01\right\rangle_{56}+c\left|10\right\rangle_{56}+d\left|11\right\rangle_{56} \right] \left|0\right\rangle_{7} \nonumber \\ &+& \frac{1}{\sqrt{2}}\left
  [-a\left|10\right\rangle_{56}-b\left|11\right\rangle_{56}+c\left|00\right\rangle_{56}+d\left|01\right\rangle_{56} \right]\left|1\right\rangle_{7}.
\end{eqnarray}
%-------------------------------------------------------------------------------------------------------------------------------------------------------------
For Charlie's outcome of 
$\left|0\right\rangle_{7}$, Bob's particles are in the state identical
to the one communicated by Alice, however, for his outcome
$\left|1\right\rangle_{7}$, Bob needs to do a $\sigma_{z}^{5}$ and
$\sigma_{x}^{5}$ operation on the $5^{th}$-particle to complete the
process successfully. Again, for all the outcomes of Alice and Charlie,
Bob can recover the message with single qubit unitary transformations,
if needed. The above processes can be generalized for the case of
$(2N+1)$ number of particles as follows. \par 
The generalized entangled basis set corresponding to Eq. (18) is 
%-----------------------Equation 22-------------------------------------------------------------------------------
\begin{eqnarray}
\left|\varphi\right\rangle_{12...2N(2N+1)}^{(1)-{(2^{2N+1})}}&=&
\frac{1}{\sqrt{2}} \left[  \left( \begin{array}{l} \left|\chi^{(1)}\right\rangle_{12...(2N-1)2N} \\
    \left|\chi^{(2)}\right\rangle_{12...(2N-1)2N}  \\
    \vdots \\ 
    \left|\chi^{(2^{2N-1}-1)} \right\rangle_{12...(2N-1)2N} \\
    \left|\chi^{(2^{2N-1})} \right\rangle_{12...(2N-1)2N} \\ \end{array} \right)
\otimes \left( \begin{array}{c}  \left|0\right\rangle_{2N+1} \\
    \left|1\right\rangle_{2N+1} \\ \end{array} \right)
\right. \nonumber \\ & & \pm
 \left. \left( \begin{array}{l} \left|\chi^{(2^{2N-1}+1)}\right\rangle_{12...(2N-1)2N}
   \\ \left|\chi^{(2^{2N-1}+2)}\right\rangle_{12...(2N-1)2N}  \\
   \vdots \\ 
    \left|\chi^{(2^{2N}-1)} \right\rangle_{12...(2N-1)2N} \\
    \left|\chi^{(2^{2N})} \right\rangle_{12...(2N-1)2N} \\ \end{array} \right)
\otimes \left( \begin{array}{c} \left|1\right\rangle_{2N+1} \\
    \left|0\right\rangle_{2N+1} \\ \end{array} \right) \right]
\end{eqnarray}
%-------------------------------------------------------------------------------------------------------------
where $\left|\chi\right\rangle^{(i)}_{12...(2N-1)2N}$'s are ordered in
the same way as in Eq. (19) and \\ $\left|\chi\right\rangle^{(i)}_{12...(2N-1)2N}$'s are
$2N$-particle generalization of Eq. (13), namely 
%----------------Equation 23--------------------------------------------------------------------------------------------------
\begin{eqnarray}
\left|\chi\right\rangle_{12..(2N-1)2N}^{(1)-(2^{2N})}&=&
\frac{1}{\sqrt{2}} \left[ \left( \begin{array}{c}\left|0\right\rangle_{1} \\  \left|1\right\rangle_{1} \\ \end{array} \right)  \otimes 
\left( \begin{array}{l}
\left| \chi^{(1)}   \right\rangle _{23 \ldots 2N}  \\
\left| \chi^{(2)}   \right\rangle _{23 \ldots 2N} \\
 \vdots   \\
\left| \chi^{(2^{2N-3}-1)}\right\rangle _{23 \ldots 2N} \\
\left| \chi^{(2^{2N-3})}\right\rangle _{23 \ldots 2N}  \\
\end{array} \right) \otimes \left( \begin{array}{c}\left|0\right\rangle_{N+1} \\ \left|1\right\rangle_{N+1} \\ 
\end{array} \right)  \right. \nonumber \\ & & 
\pm \left.
\left( \begin{array}{c} \left|1\right\rangle_{1} \\ \left|0\right\rangle_{1} \\ \end{array} \right)
\otimes \left( \begin{array}{c} \left|\chi^{(2^{2N-3}+1)}\right\rangle_{23 \ldots 2N}  \\
    \left|\chi^{(2^{2N-3}+2)}\right\rangle_{23 \ldots 2N}   \\
     \vdots \\ \left|\chi^{(2^{2N-2}-1)}\right\rangle_{23 \ldots 2N}  \\ 
  \left|\chi^{(2^{2N-2})}\right\rangle_{23 \ldots 2N} \\
\end{array} \right) \otimes \left( \begin{array}{c}\left|1\right\rangle_{N+1} \\ \left|0\right\rangle_{N+1} \\
\end{array} \right) \right].
\end{eqnarray}
%-----------------------------------------------------------------------------------------------------------------------------------------------------------------
This can be used for teleportation of
$N$-particle arbitrary information from Alice to Bob. The
$(2N+1)$-particle channel is shared between Alice and Bob such that the
first $(N+1)$ particles are with Alice and the rest $N$ particles are
with Bob. It is
important to choose the correct projection basis such that the teleportation
becomes feasible in all outcomes with only single qubit operations on
Bob's end.  For controlled teleportation the same basis set is shared between
Alice, Charlie and Bob such that the first $N$ particles are with
Alice, $(2N+1)$-th particle is with Charlie and rest $N$ particle are
with Bob. To realize successful teleportation Alice projects her
particle's on the $2N$-particle generalized basis set given by
Eq. (25). To control the process effectively Charlie measures his $(2N+1)$-th
particle in computational basis.
\subsection{Direct and controlled dense coding} 
Dense coding is concerned with the transfer of two bits of classical message between
two parties by sending only one qubit (particles) from sender to receiver provided
they share a maximally entangled pair of two qubits. In this
subsection we discuss dense coding protocol using maximally
entangled state proposed in this article.  \par
Alice and Bob must share a maximally entangled quantum channel given
by $\left|\varphi\right\rangle_{12345}^{(i)}$ such that the
qubits 1, 2, and 3 are with Alice and qubits 4 and 5 are with
Bob. Alice can locally operate her
qubits to encode the desired message, using operators from the set $(I^1, \sigma_{x}^1, \sigma_{y}^1,
\sigma_{z}^1)$, $(I^2, \sigma_{x}^2, \sigma_{y}^2,
\sigma_{z}^2)$ and $(I^3, \sigma_{x}^3, \sigma_{y}^3,
\sigma_{z}^3)$ corresponding to her three qubits 1, 2 and 3, 
respectively. Alice sends her encoded qubits to Bob who decodes
the message by an appropriate measurement on the joint five qubit
state. \par 
However, for the five-qubit quantum channel
$\left|\varphi\right\rangle^{(i)}_{12345}$, Alice will only produce 32
orthogonal states and hence can encode a 5-bit message. The operator set that is
used to prepare all the orthogonal states belonging to the entangled
set is given by $[(I^{1}I^{2}I^{3}$,
$\sigma^{1}_{z}$, $\sigma^{2}_{z}$, $\sigma^{3}_{z}$,
$\sigma^{1}_{z}\sigma^{2}_{z}$, $\sigma^{1}_{z}\sigma^{3}_{z}$, $\sigma^{2}_{z}\sigma^{3}_{z}$, 
$\sigma^{1}_{z}\sigma^{2}_{z}\sigma^{3}_{z})$, $(\sigma^{1}_{x}$,
$\sigma^{1}_{x}\sigma^{2}_{z}$, $\sigma^{1}_{x}\sigma^{3}_{z}$,
$\sigma^{1}_{x}\sigma^{2}_{z}\sigma^{3}_{z}$, $\sigma^{2}_{x}$,
$\sigma^{1}_{z}\sigma^{2}_{x}$, $\sigma^{2}_{x}\sigma^{3}_{z}$,
$\sigma^{1}_{z}\sigma^{2}_{x}\sigma^{3}_{z})$, $(\sigma^{1}_{y}$, 
$\sigma^{1}_{y}\sigma^{2}_{z}$, $\sigma^{1}_{y}\sigma^{3}_{z}$,
$\sigma^{1}_{y}\sigma^{2}_{z}\sigma^{3}_{z}$,  $\sigma^{2}_{y}$,
$\sigma^{1}_{z}\sigma^{2}_{y}$, $\sigma^{2}_{y}\sigma^{3}_{z}$,
$\sigma^{1}_{z}\sigma^{2}_{y}\sigma^{3}_{z})$, $(\sigma^{1}_{x} \sigma^{2}_{x}$,
$\sigma^{1}_{x} \sigma^{2}_{x}\sigma^{3}_{z}$, $\sigma^{1}_{x} \sigma^{2}_{y}$,
$\sigma^{1}_{x} \sigma^{2}_{y}\sigma^{3}_{z}$, $\sigma^{1}_{y} \sigma^{2}_{x}$,
$\sigma^{1}_{y} \sigma^{2}_{x}\sigma^{3}_{z}$, $\sigma^{1}_{y} \sigma^{2}_{y}$,
$\sigma^{1}_{y} \sigma^{2}_{y}\sigma^{3}_{z})]$. The capacity of dense coding
channel [64, 65] is $\chi(\rho^{AB})=log_{2}D_{A}+S(\rho^{B})-S(\rho^{AB})$
where $D_{A}$ is the dimension of Alice's system, $\rho ^B  = \mathop
{Tr}\limits_A (\rho  ^{AB} )$ is
reduced density matrix of Bob's system with respect to Alice's system
and $S(\rho ) =  - Tr(\rho \log _2 \rho )$
is the von Neumann entropy. For the entangled channel
$\left|\varphi\right\rangle_{12345}^{(i)}$,  $D_{A}=2^{3}$, $S(\rho^{B})=2$, and
$S(\rho^{AB})=0$ which shows $\chi(\rho^{AB})=5$ and thus maximizes
the channel capacity. It has been shown earlier that by using a maximally entangled five-qubit GHZ state or a
generalized five-qubit W state as a quantum
channel, Alice can only send 4-bit information.  \par
Quantum dense coding can also be realized
involving a controller between Alice and Bob. By doing so, we compare
the transfer efficiency of the channel proposed by us to that of
others discussed in section 2. We do this by using three different states of maximally entangled five-qubits
%--------------------------Equation 24----------------------------------------------------------------------------------------------------------------------------
\begin{eqnarray*} (i)
\left|\Psi\right\rangle^{(1)}_{12345}&=&\frac{1}{2}\left[
     \left|00000\right\rangle + \left|10101\right\rangle +
     \left|01110\right\rangle - \left|11011\right\rangle \right
   ]_{12345}.
\end{eqnarray*}  \begin{eqnarray*} (ii)
\left|\Phi\right\rangle^{(1)}_{12345}&=&\frac{1}{2}\left[
     \left|00000\right\rangle + \left|10101\right\rangle +
     \left|01110\right\rangle + \left|11011\right\rangle \right
   ]_{12345}. 
\end{eqnarray*}
\begin{eqnarray} (iii) \left|\varphi\right\rangle^{(10)}_{12345}&=&\frac{1}{2\sqrt{2}}\left[
     \left|00000\right\rangle - \left|00101\right\rangle +
     \left|11100\right\rangle + \left|11001\right\rangle
   \right. \nonumber \\ &+& \left. \left|01111\right\rangle- \left|01010\right\rangle +
     \left|10011\right\rangle + \left|10110\right\rangle \right
   ]_{12345}
\end{eqnarray} 
%-------------------------------------------------------------------------------------------------------------------------------------------------------------
given by Eq. (16), Eq. (17) and Eq. (18), respectively.  \par
The quantum channel $\left|\Psi\right\rangle_{12345}$ is shared by
three users Alice (1 and 2), Charlie (3) and Bob (4 and 5). Charlie
performs a von Neumann measurement on
his share of qubit in the new
basis $(\left|x_{1}\right\rangle_{3}, \left|x_{2}\right\rangle_{3})$
such that
%------------------------------Equation 25---------------------------------------------------------------------------------------------------------------------
\begin{eqnarray}
\left|0\right\rangle_{3} & = & \cos{\theta} \left|x_{1}\right\rangle_{3} +
\sin{\theta} \left|x_{2}\right\rangle_{3} {\rm ~ ~ ~ and} \nonumber \\ 
\left|1\right\rangle_{3} & = & \sin{\theta} \left|x_{1}\right\rangle_{3} -
\cos{\theta} \left|x_{2}\right\rangle_{3}
\end{eqnarray} 
%---------------------------------------------------------------------------------------------------------------------------------------------------------------
where 
$\theta = \theta$ if $0\leq \theta \leq (\pi)/4$. For $\theta$ in the
range $(\pi)/4\leq \theta \leq
(\pi)/2$, replace $\theta$ in Eq. [27] by $({90}^{o}-\theta)$.
In this basis, $\left|\Psi\right\rangle_{12345}$ is given by 
%------------------------------Equation 26---------------------------------------------------------------------------------------------------------------------
\begin{eqnarray}
\lefteqn{\left|\Psi\right\rangle_{12345} =} & & \nonumber \\ & & 
\frac{\left|x_{1}\right\rangle_{3}}{2}\left[
  \cos{\theta}\left|0000\right\rangle+\sin{\theta}\left|1001\right\rangle+\sin{\theta}\left|0110\right\rangle-\cos{\theta}\left|1111\right\rangle \right]_{1245} \nonumber \\ & + & \frac{\left|x_{2}\right\rangle_{3}}{2}\left[
  \sin{\theta}\left|0000\right\rangle-\cos{\theta}\left|1001\right\rangle-\cos{\theta}\left|0110\right\rangle-\sin{\theta}\left|1111\right\rangle \right]_{1245}.
\end{eqnarray}
%--------------------------------------------------------------------------------------------------------------------------------------------------------------
In the special case of  $\theta=\pi/4$, the four
qubits are in maximally entangled four-qubit state and
Alice can send 4 bits of message to Bob by first encoding the message
and then sending 2 qubits of hers to him. However, Charlie
can do measurement for any $\theta$. \\ 
(A) If Charlie's result is $\left|x_{1}\right\rangle_{3}$ and he
communicates his measurement result to Alice then the dense coding
process is given schematically in Fig. (3). 
Alice introduces an auxiliary qubit $\left|0\right\rangle_{aux}$ and does a
joint unitary operation $U_{12aux}$ on her qubits 1, 2 and auxiliary qubit
(in the computational basis for qubits 1, 2 and 3). The unitary operation is 
%------------------------------Equation 27---------------------------------------------------------------------------------------------------------------------
\begin{eqnarray} 
\lefteqn{U_{12aux} =} & & \nonumber \\ & & 
\left( \begin{array}{cccccccc}
\frac{\sin{\theta}}{\cos{\theta}} &
\sqrt{1-\frac{\sin^2{\theta}}{\cos^2{\theta}}} & 0 & 0 & 0 & 0 & 0 & 0
\\ \sqrt{1-\frac{\sin^2{\theta}}{\cos^2{\theta}}} & \frac{-\sin{\theta}}{\cos{\theta}} & 0 & 0 & 0 & 0 & 0 & 0 \\ 0 &
0 & 1 & 0 & 0 & 0 & 0 & 0 \\ 0 & 0 & 0 & 1 & 0 & 0 & 0 & 0 \\ 0 & 0 & 0 & 0 & 1 & 0 & 0 & 0 \\ 0 & 0 & 0 & 0 & 0 & 1 & 0 & 0 \\ 0 &
0 & 0 & 0 & 0 & 0 & \frac{\sin{\theta}}{\cos{\theta}} &
\sqrt{1-\frac{\sin^2{\theta}}{\cos^2{\theta}}} \\ 0 & 0 & 0 & 0 & 0 &
0 & \sqrt{1-\frac{\sin^2{\theta}}{\cos^2{\theta}}} & \frac{-\sin{\theta}}{\cos{\theta}} \end{array}
\right). \nonumber \\ & &
\end{eqnarray} 
%--------------------------------------------------------------------------------------------------------------------------------------------------------------
If Charlie's measurement outcome is $\left|x_{2}\right\rangle_{3}$,
the unitary operation that Alice will use is  $U^{'}_{12aux}$ where 
%------------------------------Equation 28---------------------------------------------------------------------------------------------------------------------
\begin{eqnarray} 
\lefteqn{U^{'}_{12aux} =} & & \nonumber \\ & & 
\left( \begin{array}{cccccccc} 
1 &
0 & 0 & 0 & 0 & 0 & 0 & 0
\\0 & 1 & 0 & 0 & 0 & 0 & 0 & 0 \\ 0 &
0 & \frac{\sin{\theta}}{\cos{\theta}} &
\sqrt{1-\frac{\sin^2{\theta}}{\cos^2{\theta}}} & 0 & 0 & 0 & 0 \\ 0 &
0 &  \sqrt{1-\frac{\sin^2{\theta}}{\cos^2{\theta}}} &
\frac{-\sin{\theta}}{\cos{\theta}} & 0 & 0 & 0 & 0 \\ 0 & 0 & 0 & 0 & \frac{\sin{\theta}}{\cos{\theta}} &
\sqrt{1-\frac{\sin^2{\theta}}{\cos^2{\theta}}} & 0 & 0 \\  0 & 0 & 0 &
0 & \sqrt{1-\frac{\sin^2{\theta}}{\cos^2{\theta}}} &
\frac{-\sin{\theta}}{\cos{\theta}} & 0 & 0 \\ 0 & 0 & 0 &
0 & 0 & 0 & 1 & 0 \\ 0 &
0 & 0 & 0 & 0 & 0 & 0 &
1 \\ \end{array}
\right).  \nonumber \\ & &
\end{eqnarray}
%----------------------------------------------------------------------------------------------------------------------------------------------------------------
If Alice makes a measurement on auxiliary qubit and gets the measurement
result 
$\left|0\right\rangle_{aux}$ (with the 
probability $2\sin^{2}{\theta}$), she knows that the joint state of four
qubits is in
a maximally entangled four-qubit state; however, the measurement
result $\left|1\right\rangle_{aux}$ (with the 
probability $\cos{2\theta}$), will confirm that the
four-qubits are in GHZ state. In the first case, Alice
can encode her message using 16 binary operators from the set $(I^{1},
\sigma_{x}^{1}, \sigma_{y}^{1}, \sigma_{z}^{1})$ and  $(I^{2},
\sigma_{x}^{2}, \sigma_{y}^{2}, \sigma_{z}^{2})$ corresponding to her
two qubits 1 and 2. She sends her two qubits to Bob who decodes the
message by doing a joint measurement on four-qubits based on the  entangled state Alice has prepared and thus decodes the
original message. However, doing a joint measurement on four-qubits
to discriminate 16 orthogonal states 
is experimentally challenging. Hence the following: \\
(i) Bob does a joint unitary operation on Alice's $1^{st}$ qubit and
his $5^{th}$ qubit given as $U_{A_{1}B_{5}} = \frac{1}{\sqrt{2}}\left(
    \begin{array}{cccc} 1 & 0 & 0 & 1
\\ 0 & 1 & 1 & 0 \\ 0 & 1 & -1 & 0  \\ 1 & 0 & 0 & -1  \end{array}
\right)$ which evolves the four-qubit state(s) into a four-qubit GHZ
state(s), measures the joint state in the four-qubit GHZ basis and
recovers the encoded message. \\
(ii) Bob applies two C-NOT operations on the state of four qubits keeping Alice's qubits 1 and 2 as
controls and his qubits 5 and 4 as targets, respectively which leads to a state of direct
products of four qubits; Alice's two qubits are in an
entangled state and Bob's two qubits are in a computational basis
state. By doing these operations
Bob ensures that he differentiates between the subsets of Alice's operations,
i.e. if he measures his two qubits in computational basis states and
finds $\left|00\right\rangle_{45}$, he knows that the operation Alice has used to encode the
message belongs to the subset $(I^{1}I^{1}, \sigma_{z}^{1},
\sigma_{z}^{2}, \sigma_{z}^{1}\otimes
\sigma_{z}^{2})$. Similarly other measurements 
$\left|01\right\rangle_{45}$, $\left|10\right\rangle_{45}$ and 
$\left|11\right\rangle_{45}$ belong to the subsets $ (\sigma_{x}^{1},
\sigma_{y}^{1}, \sigma_{x}^{1}\otimes
\sigma_{z}^{2},  \sigma_{y}^{1}\otimes
\sigma_{z}^{2})$, $ (\sigma_{x}^{2},
\sigma_{y}^{2}, \sigma_{x}^{2}\otimes
\sigma_{z}^{1},  \sigma_{y}^{2}\otimes
\sigma_{z}^{1})$ and  $(\sigma_{x}^{1}\otimes
\sigma_{x}^{2},  \sigma_{x}^{1}\otimes
\sigma_{y}^{2}, \sigma_{y}^{1},
\sigma_{x}^{2}, \sigma_{y}^{1}\otimes
\sigma_{y}^{2})$, respectively. Bob uses Bell basis to measure
Alice's qubits  (which are just a local transformation away from
Bell states) and decodes the 4-bit classical message with relative ease. \par
In the other case, where Alice's measurement result yields 
$\left|1\right\rangle_{aux}$, she can still send 3-bit classical
message to Bob. Therefore on the average  
\begin{eqnarray}
%----------------------------Equation 29-----------------------------------------------------------------------------------------------------------------
I_{C-A}^{(1)} &=& 2\sin^{2}{\theta}+3
\end{eqnarray}
%-----------------------------------------------------------------------------------------------------------------------------------------------------------
-bit classical message is transferred where the suffix C-A denotes that Charlie informs his measurement
result to Alice only. \\
(B)  If Charlie sends his measurement result to Bob and not to Alice, then the process
for sending the information is pictorially represented in Fig. (4). 
Alice does local operations on her qubits and sends
them to Bob who,  cannot however, do a joint measurement to discriminate
all the state of four-qubits as they may or may not be orthogonal. We take Alice's first
operational  subset $(I^{1}I^{1}, \sigma_{z}^{1},
\sigma_{z}^{2}, \sigma_{z}^{1}\otimes
\sigma_{z}^{2})$, as an example, such that the four-qubit state
immediately after Alice's operation is $\left|\psi\right\rangle_{1245}=\frac{1}{\sqrt{2}}\left[
  \cos{\theta}\left|0000\right\rangle \pm   \sin{\theta}
  \left|1001\right\rangle\pm\sin{\theta}
  \left|0110\right\rangle\ \mp   \cos{\theta}\left|1111\right\rangle \right]_{1245}$ and where -ve signs are always in
  odd numbers.  
After receiving her qubits Bob does two C-NOT operations as described
in the case (A). Alice's qubits are  partially entangled, whereas Bob's
qubits are in computational basis state such that $\left|\psi\right\rangle_{1245}=\frac{1}{\sqrt{2}}\left[
  \cos{\theta}\left|00\right\rangle \pm   \sin{\theta}
  \left|01\right\rangle \right. \nonumber \\ \pm
    \sin{\theta} \left.\left|10\right\rangle \mp   \cos{\theta}
    \left|11\right\rangle \right ]_{12} \otimes
  \left|00\right\rangle_{45}$. By measuring his qubits in the 
  computational basis, Bob gets Alice's operational subset on her two
  qubits, thus he needs to differentiate between four states of
  two-qubit system related to each operational subset. For this Bob
  introduces an auxiliary qubit in state $\left|0\right\rangle_{aux}$
  and performs a joint unitary operation given by Eq. (29) on three
  qubits (1, 2 and aux) in the computational basis of $12aux$, such
  that 
%-------------------------------Equation 30----------------------------------------------------------------------------------------------------------------------
\begin{eqnarray}
\left|\psi'\right\rangle_{12aux} &=& (U_{12aux})
\left|\psi\right\rangle_{12aux} \nonumber \\ &=&
{\sqrt{2}} \, \, {\sin{\theta}}.\frac{1}{2}\left[
  \left|00\right\rangle \pm \left|01\right\rangle \pm
  \left|10\right\rangle \mp \left|11\right\rangle
  \right]_{12}\otimes \left|0\right\rangle_{aux} \nonumber \\ & + &
  \cos{\theta}.\sqrt{1-\frac{\sin^2{\theta}}{\cos^2{\theta}}}.\frac{1}{\sqrt{2}}\left[ \left|00\right\rangle \mp \left|11\right\rangle
  \right]_{12}\otimes \left|1\right\rangle_{aux}.
\end{eqnarray} 
%----------------------------------------------------------------------------------------------------------------------------------------------------------------
The average information transfer from
Alice to Bob will then be  
%-------------------------------Equation 31--------------------------------------------------------------------------------------
\begin{equation}
I_{C-B}^{(2)} = 2\sin^{2}{\theta}+3
\end{equation}
%-------------------------------------------------------------------------------------------------------------------------------------------------------
where the suffix C-B denotes that Charlie informs his measurement
result to Bob only. It is clear that in both the cases [(A) and (B)] the amount of
information transfer, on average, between Alice and Bob is the same. 
However, in the case of a five-qubit GHZ state the
amount of information transferred in controlled manner is
$2\sin^{2}{\theta}+2$. \par 
A similar calculation for information transfer
between Alice and Bob using the quantum channel proposed,
$\left|\varphi\right\rangle^{(i)}_{12345}$, shows that irrespective of the
measurement basis used by Charlie, Alice is always able to send 4-bit
information to Bob using her 2-qubits. Even if Charlie does not inform Alice
about the measurement basis used, she is able to send maximum
information to Bob using her two qubits such that the information
transfer between Alice and Bob is independent of
the value of analyzer angle  $\theta$. At the
same time, the information transfer between Alice and Bob using
$\left|\Phi\right\rangle_{12345}^{(1)}$ is the same as one obtains in
the case
of $\left|\Psi\right\rangle_{12345}^{(1)}$. It is
seen that all the five-qubit states used for
controlled dense coding possess maximum correlation between the
particles, however, the amount of information transfer using different
quantum channels is not the same.  The entangled set proposed here is advantageous in terms of average information transfer between the sender and the receiver as compared to other five-qubit entangled sets. Thus the
representation of a quantum channel used in a communication network
and distribution of qubits between different users is an
important factor in information processing.  
A graphical comparison is made in Fig. (5) to compare the efficiency of
genuinely entangled five-qubit states discussed here and the five-qubit GHZ
states in terms of average information transferred during the process
and the analyzer angle $\theta$. \par
For generalized dense coding the $(2N+1)$-qubit quantum channel is shared between
Alice and Bob such that $(N+1)$  qubits $(1$ to $N+1)$ are with Alice and
rest of the qubits are with Bob. By locally manipulating her
qubits, Alice encodes her message and sends her qubits to Bob who, in
turn, does the required measurements involving all the qubits and
decodes the message. In principle Alice can prepare $2^{2N+1}$
orthogonal basis states and hence $2^{2N+1}$  distinguishable messages
for Bob. Thus by using the generalized entangled channels Alice can
send $(2N+1)$-bit information to Bob. 
Alternatively, the $(2N+1)$-qubit state is shared between the three users
in communication protocol such that $N$ qubits  $(1$ to $N)$ are with Alice,
one qubit is with Charlie [$(N+1)$-th] and the rest $N$ qubits are with
Bob $(N + 2 \to 2N + 1)$. Charlie measures his qubit in the basis given by
Eq. (27) 
and either sends his measurement results to Alice or to Bob. There are
two instances: \\ 
(A) If Charlie sends his measurement results to Alice, she introduces
an auxiliary qubit and does a combined unitary transformation on her
$N$ qubits and the auxiliary qubit. After this she measures the state of
auxiliary qubit, encodes her message and sends her qubits to Bob who does a joint measurement on $2N$ qubits and decodes the
message. In practice it is really difficult to discriminate
multi-qubit states experimentally. As an alternate, Bob can do $N$
C-NOT operations keeping Alice's qubits $1, 2, 3,.....,N$ as controls and
his qubits $(2N+1), 2N, (2N-1),.....,(N+2)$, respectively as targets and
measure last $(2N-2)$ qubits in a computational basis and first two
qubits in Bell basis (which are generally a local transformation away
from Bell states) and recover the message. \\
(B) If Charlie sends his measurement results to Bob, Alice encodes her
message and sends her qubits to Bob
who applies $N$ C-NOT operations as discussed
above, measures last $(2N-2)$ qubits in a computational basis and then
introduces an auxiliary qubit so that he can discriminate Alice's
operation by doing joint unitary transformation on three qubits. After
the transformation, Bob measures the state of the auxiliary qubit and
discriminates Alice's operations to decode the message. \par
Although the average information transferred in both the cases is
$[2\sin^2{\theta}+(2N-1)]$, the case where Charlie sends his
measurement results to Bob is more appealing as the joint
transformation that Bob needs to do involves only 3 qubits, however, the
one which Alice does involves $(N+1)$ qubits. In the case of $(2N+1)$-qubit
generalization of $\left|\varphi\right\rangle_{12345}^{(i)}$, Alice is able to
send a $2N$-bit information to Bob independent of Charlie's measurement
basis. 
\section{CONCLUSIONS}
We have given a criterion to assess the degree of entanglement between
qubits/spin-1/2 particles
using statistical correlation coefficients as a measure of entanglement. Ursell-Mayer
type correlation functions have been suggested to calculate the correlations
between multiple particles
which are extensions to the two-particle functions. The use
of Ursell-Mayer type correlation functions to calculate the entanglement between
multi-particles is an attempt to generalize the definition of degree of entanglement in multi-particle systems irrespective of the number of particles involved. The criterion is shown to be unique in characterizing different entangled systems in different families. It has been shown that the local transformations between two-states can be established by visual examination of the non-zero correlation coefficients
associated with the systems under study. The criterion developed here
is compared with the existing entanglement norms for two and three
particle maximally and non-maximally entangled systems and found to be
in excellent match.\par
We have proposed a maximally and genuinely entangled five-particle
quantum channel and described an efficient theoretical approach for direct quantum teleportation of
multi-particle information. The process discussed here overcomes the difficulty of getting null
results in half of the projections when dealing with odd number of
particles comprising a quantum channel. Teleportation using a
controller has also been shown to be effective using appropriate projection basis. Physical realization of states and quantum teleportation protocols are analyzed using standard quantum gates and circuit diagrams. Quantum dense coding using the
state proposed has been shown to be optimal. For controlled dense
coding process it has been observed that unlike cluster and $\left|\chi\right\rangle$ type of states where amount of information transfer depends on the analyzer angle used by the controller,  the information transfer using the state proposed is
always maximum irrespective of the measurement basis used by the controller.
A comparison between average information transferred in
the case of the state proposed here and other five-qubit states including GHZ
state has been made
through a plot at various analyzer angles to have more insight into
the entanglement properties and representation of quantum channel used. It has been observed that for higher number of qubits when a controller is involved, it is desirable
that measurement results be transmitted to the receiver and not to the sender. 
\section*{Acknowledgements}
AK is grateful to IIT Madras for a graduate fellowship. MSK  would like
to acknowledge IIT Madras for research funds. 

\section*{\bf Appendix A: Correlation coefficients for five-particles}
The correlation coefficients
$C_{\alpha\beta\gamma\delta\kappa}^{12345}$ for five-particle systems
are given by 
%-----------------------------------------Equation ----------------------------------------------------------------------
\begin{eqnarray*} 
\lefteqn{C^{12345}_{\alpha \beta \gamma \delta \kappa} =} & & \nonumber \\ & & 
\left\langle \sigma^1_{\alpha} \sigma^2_{\beta}
  \sigma^3_{\gamma}\sigma^4_{\delta} \sigma^5_{\kappa}
  \right\rangle-\left \langle \sigma^{1}_{\alpha}\right \rangle \left[C^{2345}_{\beta\gamma\delta\kappa}\right]
-\left \langle \sigma^{2}_{\beta}\right \rangle \left[C^{1345}_{\alpha\gamma\delta\kappa}\right]
-\left \langle \sigma^{3}_{\gamma}\right \rangle
\left[C^{1245}_{\alpha\beta\delta\kappa}\right] \nonumber \\ &-& 
\left \langle \sigma^{4}_{\delta}\right \rangle \left[C^{1235}_{\alpha\beta\gamma\kappa}\right]
-\left \langle \sigma^{5}_{\kappa}\right \rangle \left[C^{1234}_{\alpha\beta\gamma\delta}\right]
-\left \langle \sigma^1_{\alpha} \sigma^2_{\beta} \right \rangle 
\left \langle \sigma^3_{\gamma} \sigma^4_{\delta}
  \sigma^5_{\kappa}\right \rangle \nonumber \\ &-&  \left \langle \sigma^1_{\alpha} \sigma^3_{\gamma} \right \rangle 
\left \langle \sigma^2_{\beta} \sigma^4_{\delta}
  \sigma^5_{\kappa}\right \rangle -
\left \langle \sigma^1_{\alpha} \sigma^4_{\delta} \right \rangle 
\left \langle \sigma^2_{\beta} \sigma^3_{\gamma} \sigma^5_{\kappa}\right \rangle -
\left \langle \sigma^1_{\alpha} \sigma^5_{\kappa} \right \rangle 
\left \langle \sigma^2_{\beta} \sigma^3_{\gamma}
  \sigma^4_{\delta}\right \rangle \nonumber \\ &-&  \left \langle \sigma^2_{\beta} \sigma^3_{\gamma} \right \rangle 
\left \langle \sigma^1_{\alpha} \sigma^4_{\delta}
  \sigma^5_{\kappa}\right \rangle- \left \langle \sigma^2_{\beta} \sigma^4_{\delta} \right \rangle 
\left \langle \sigma^1_{\alpha} \sigma^3_{\gamma}
  \sigma^5_{\kappa}\right \rangle - \left \langle \sigma^2_{\beta} \sigma^5_{\kappa} \right \rangle 
\left \langle \sigma^1_{\alpha} \sigma^3_{\gamma} \sigma^4_{\delta}
\right \rangle \nonumber \\ &-& \left \langle \sigma^3_{\gamma} \sigma^4_{\delta} \right \rangle 
\left \langle \sigma^1_{\alpha} \sigma^2_{\beta}
  \sigma^5_{\kappa}\right \rangle -  \left \langle \sigma^3_{\gamma} \sigma^5_{\kappa} \right \rangle 
\left \langle \sigma^1_{\alpha} \sigma^2_{\beta} \sigma^4_{\delta}\right \rangle  - \left \langle \sigma^4_{\delta} \sigma^5_{\kappa} \right \rangle 
\left \langle \sigma^1_{\alpha} \sigma^2_{\beta}
  \sigma^3_{\gamma}\right \rangle \nonumber \\ &+& \left\langle
  \sigma^1_{\alpha} \right\rangle \left\langle \sigma^2_{\beta}
  \sigma^3_{\gamma}\right\rangle \left\langle \sigma^4_{\delta}
  \sigma^5_{\kappa}\right\rangle +  \left\langle
  \sigma^1_{\alpha} \right\rangle \left\langle \sigma^2_{\beta}
  \sigma^4_{\delta}\right\rangle \left\langle \sigma^3_{\gamma}
  \sigma^5_{\kappa}\right\rangle +  \left\langle
  \sigma^1_{\alpha} \right\rangle \left\langle \sigma^2_{\beta}
  \sigma^5_{\kappa}\right\rangle \left\langle \sigma^3_{\gamma}
  \sigma^4_{\delta}\right\rangle \nonumber \\ &+&  \left\langle
  \sigma^2_{\beta} \right\rangle \left\langle \sigma^1_{\alpha}
  \sigma^3_{\gamma}\right\rangle \left\langle \sigma^4_{\delta}
  \sigma^5_{\kappa}\right\rangle + \left\langle
  \sigma^2_{\beta} \right\rangle \left\langle \sigma^1_{\alpha}
  \sigma^4_{\delta}\right\rangle \left\langle \sigma^3_{\gamma}
  \sigma^5_{\kappa}\right\rangle + \left\langle
  \sigma^2_{\beta} \right\rangle \left\langle \sigma^1_{\alpha}
  \sigma^5_{\kappa}\right\rangle \left\langle \sigma^3_{\gamma}
  \sigma^4_{\delta}\right\rangle \nonumber \\ &+&   \left\langle
  \sigma^3_{\gamma} \right\rangle \left\langle \sigma^1_{\alpha}
  \sigma^2_{\beta}\right\rangle \left\langle \sigma^4_{\delta}
  \sigma^5_{\kappa}\right\rangle + \left\langle
  \sigma^3_{\gamma} \right\rangle \left\langle \sigma^1_{\alpha}
  \sigma^4_{\delta}\right\rangle \left\langle \sigma^2_{\beta} \sigma^5_{\kappa}\right\rangle+  \left\langle
  \sigma^3_{\gamma} \right\rangle \left\langle \sigma^1_{\alpha}
  \sigma^5_{\kappa}\right\rangle \left\langle \sigma^2_{\beta}
  \sigma^4_{\delta}\right\rangle \nonumber \\ &+&   \left\langle
  \sigma^4_{\delta} \right\rangle \left\langle \sigma^1_{\alpha}
  \sigma^2_{\beta}\right\rangle \left\langle \sigma^3_{\gamma}
  \sigma^5_{\kappa}\right\rangle+ \left\langle
  \sigma^4_{\delta} \right\rangle \left\langle \sigma^1_{\alpha}
  \sigma^3_{\gamma}\right\rangle \left\langle \sigma^2_{\beta}
  \sigma^5_{\kappa}\right\rangle +  \left\langle
  \sigma^4_{\delta} \right\rangle \left\langle \sigma^1_{\alpha}
  \sigma^5_{\kappa}\right\rangle \left\langle \sigma^2_{\beta}
  \sigma^3_{\gamma}\right\rangle \nonumber \\ &+&   \left\langle
  \sigma^5_{\gamma} \right\rangle \left\langle \sigma^1_{\alpha}
  \sigma^2_{\beta}\right\rangle \left\langle \sigma^3_{\gamma}
  \sigma^4_{\delta}\right\rangle+  \left\langle
  \sigma^5_{\gamma} \right\rangle \left\langle \sigma^1_{\alpha}
  \sigma^3_{\gamma}\right\rangle \left\langle \sigma^2_{\beta}
  \sigma^4_{\delta}\right\rangle + \left\langle
  \sigma^5_{\gamma} \right\rangle \left\langle \sigma^1_{\alpha}
  \sigma^4_{\delta}\right\rangle \left\langle \sigma^2_{\beta}
  \sigma^3_{\gamma}\right\rangle  \nonumber \\ &-&  6 \left\langle
  \sigma^1_{\alpha}\right\rangle  \left\langle
  \sigma^2_{\beta}\right\rangle \left\langle
  \sigma^3_{\gamma}\right\rangle  \left\langle
  \sigma^4_{\delta}\right\rangle  \left\langle
  \sigma^5_{\kappa}\right\rangle.
\end{eqnarray*}
%--------------------------------------------------------------------------------------------------------------------------

\newpage
\begin{center}
List of Figures
\end{center}

\begin{enumerate}
\item Quantum circuit to prepare the set of states
      $\left|\varphi\right\rangle_{12345}^{(i)}$ represented by
      Eq. (18). 
\item Quantum network to teleport an arbitrary two qubit state
      $\left|\phi\right\rangle_{12}=\left[\left|00\right\rangle_{12}+\left|01\right\rangle_{12}\right. \nonumber \\ +\left.\left|10\right\rangle_{12}+\left|11\right\rangle_{12}\right]$ using $\left|\varphi\right\rangle_{34567}^{(10)}$ as quantum channel. 
\item Controlled dense coding of five-qubit state
$\left|\Psi\right\rangle_{12345}^{(1)}$ with controller-sender
interface. 
\item Controlled dense coding of five-qubit state
$\left|\Psi\right\rangle_{12345}^{(1)}$ with controller-receiver 
interface.
\item Comparison of the efficiency of information transfer between states
given by Eq. (26) and the five qubit GHZ state.
\end{enumerate}

\newpage

%-------------------figure 1-------------------------------------------------------------
\begin{center}
Figure 1
\end{center}
\begin{figure}[!htb]
\vspace*{-0.9 in}
\includegraphics [width=4.5in] {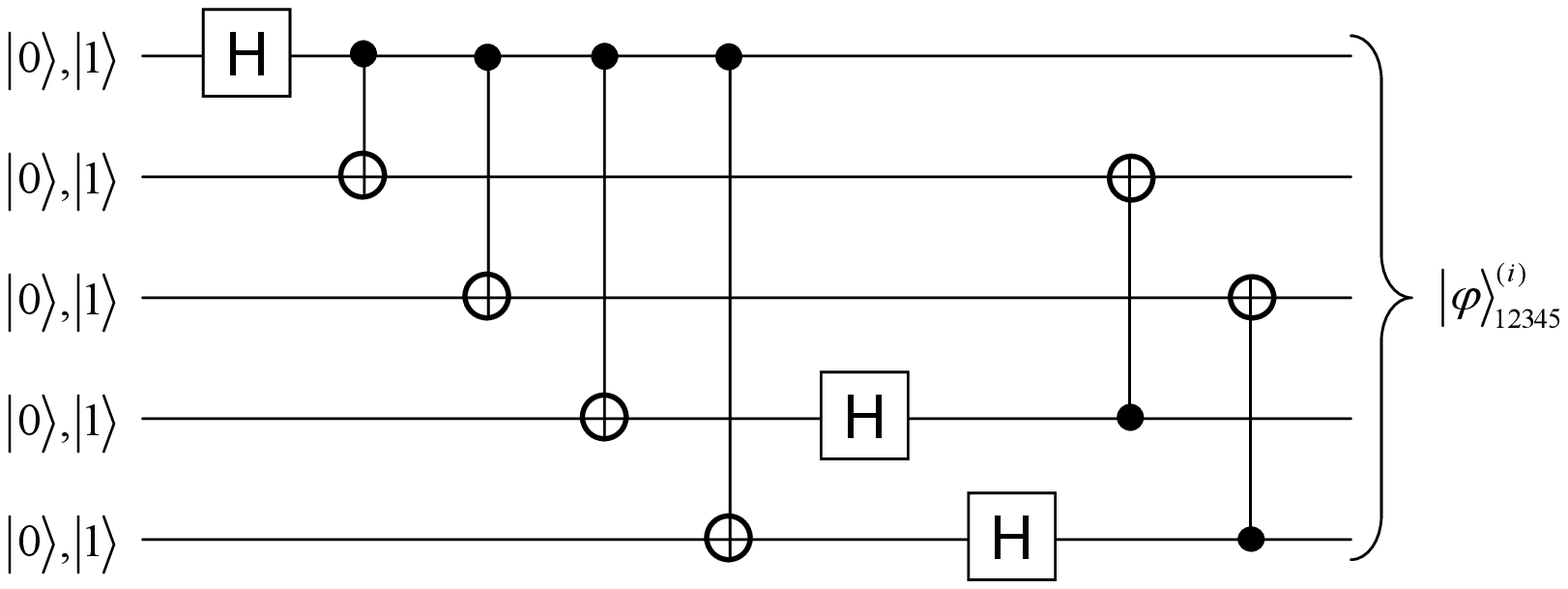}
\vspace*{-1.2 in}
\caption{Quantum circuit to prepare the set of states
      $\left|\varphi\right\rangle_{12345}^{(i)}$.}
\end{figure}
%--------------------------------------------------------------------------------------

%-------------------figure 2-------------------------------------------------------------
\newpage
\begin{center}
Figure 2
\end{center}
\begin{figure}[!htb]
\vspace*{-0.6 in}
\includegraphics [width=4.5in] {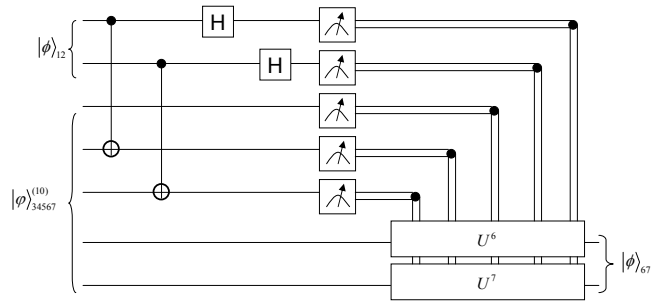}
\vspace*{-1.0 in}
\caption{Quantum network to teleport an arbitrary two particle state
  through $\left|\varphi\right\rangle_{34567}^{(1)}$.}
\end{figure}
%--------------------------------------------------------------------------------------

%-------------------figure 3-------------------------------------------------------------
\newpage
\begin{center}
Figure 3
\end{center}
\begin{center}
\end{center}
\begin{figure}[!htb]
\vspace*{-0.2 in}
\includegraphics [width=4.9in] {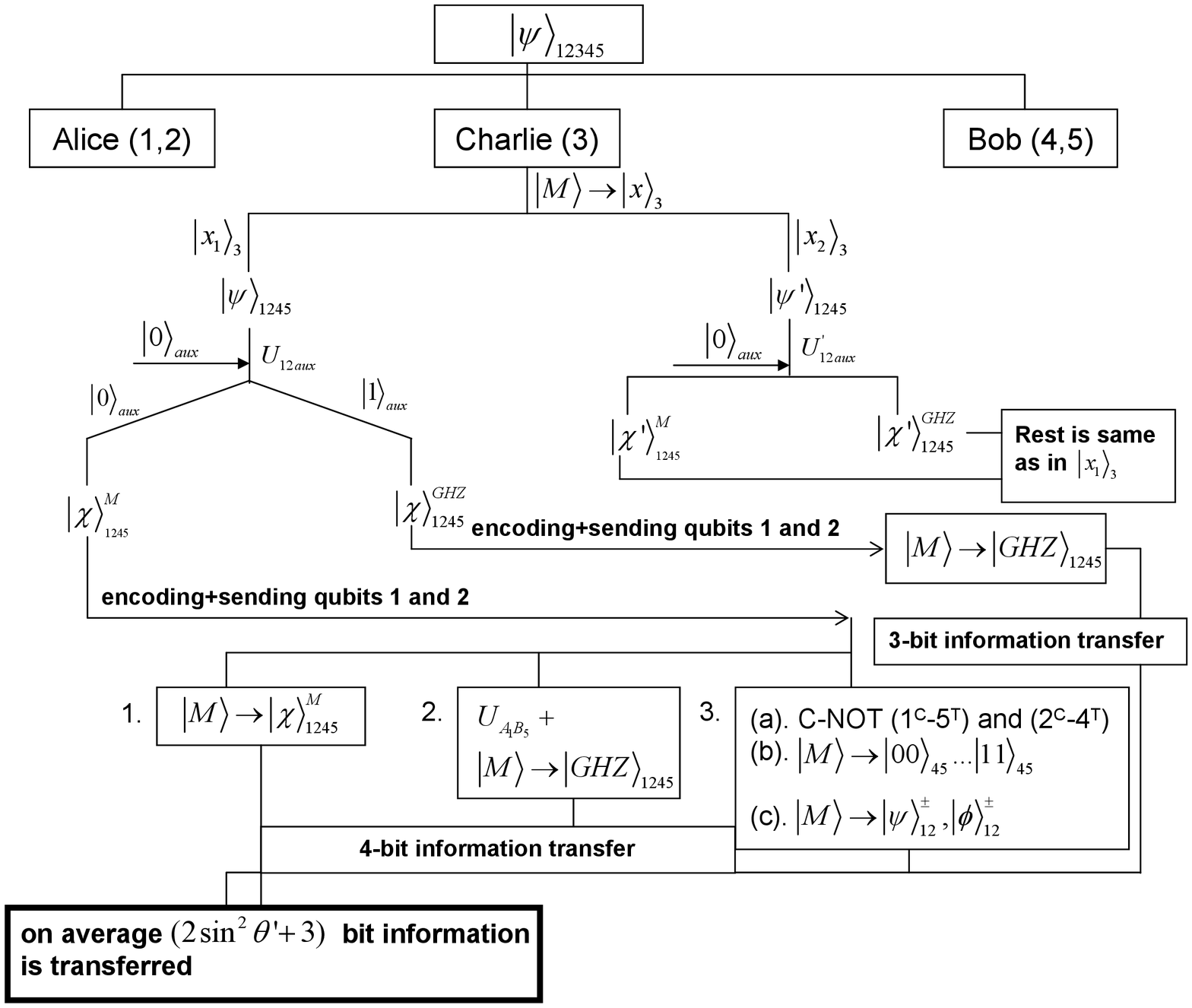}
\vspace*{-0.0 in}
\caption{Schematic representation of controlled dense coding process
         with controller-sender interface.}
\end{figure}
%--------------------------------------------------------------------------------------

%-------------------figure 4-------------------------------------------------------------
%
\newpage
\begin{center}
Figure 4 
\end{center}
\begin{center}
\end{center}
\begin{figure}[!htb]
\vspace*{-0.2 in}
\includegraphics [width=5.0in] {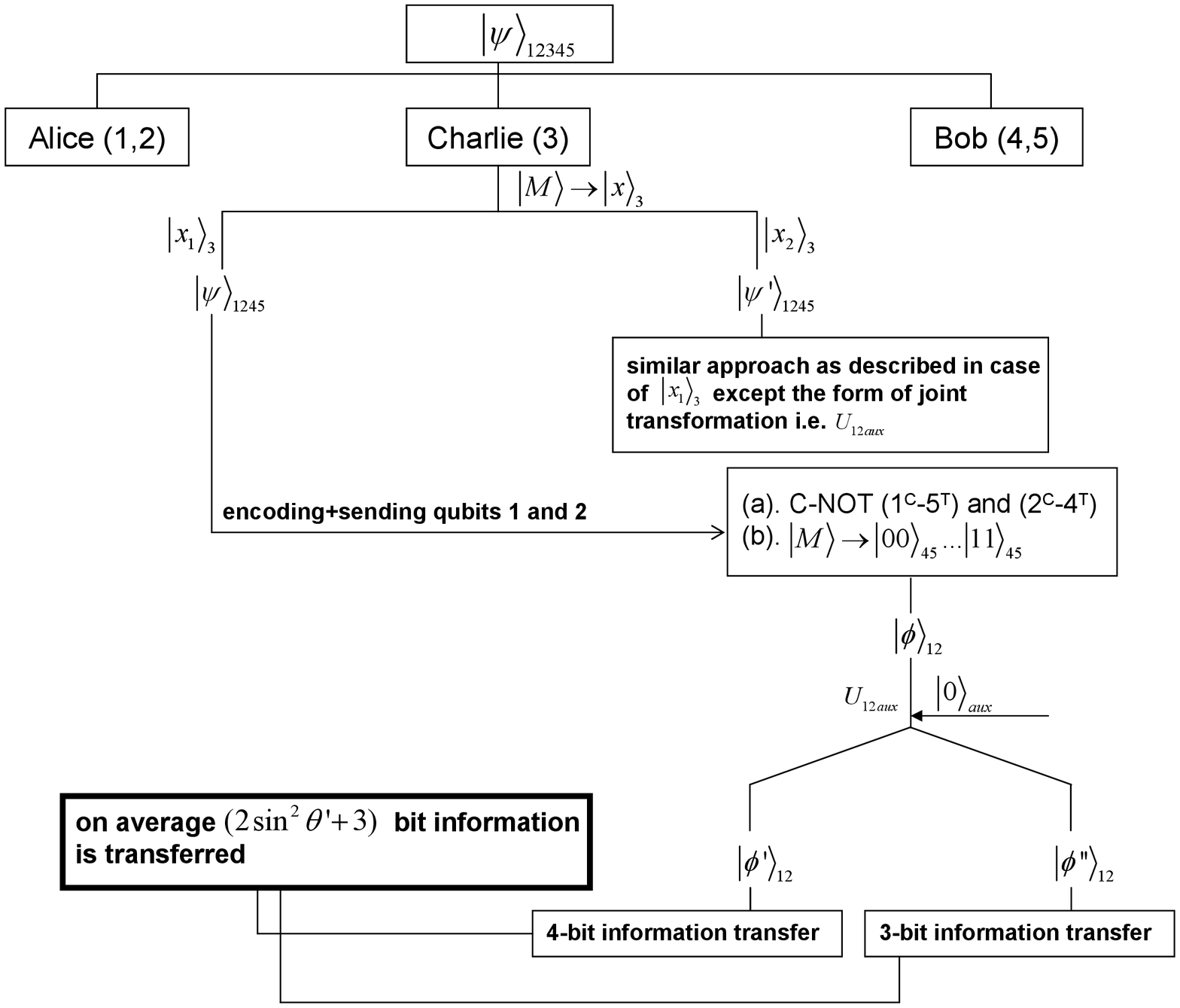}
\vspace*{-0.0 in}
\caption{Schematic representation of controlled dense coding process
         with controller-receiver interface.}
\end{figure}
%
%
%--------------------------------------------------------------------------------------

%-------------------figure 5-------------------------------------------------------------
%
\newpage
\begin{center}
Figure 5
\end{center}
\begin{center}
\end{center}
\begin{figure}[!htb]
\vspace*{-1.0 in}
\includegraphics [width=5.0in] {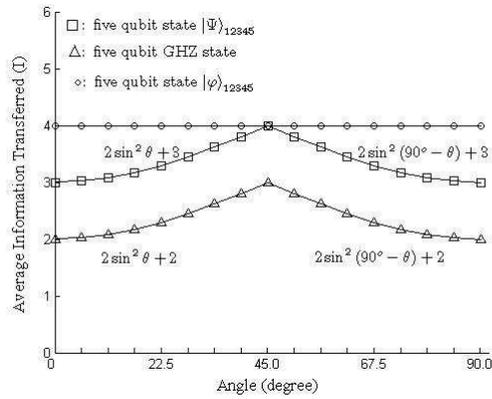}
\vspace*{-0.8 in}
\caption{Comparison of the efficiency of information transfer between states
given by Eq. (26) and the five qubit GHZ state.}
\end{figure}
%
%
%--------------------------------------------------------------------------------------

\end{document}